\documentclass[12pt]{iopart}
\RequirePackage[l2tabu,orthodox]{nag} 
\pdfoutput=1

\usepackage{iopams}

\usepackage[T1]{fontenc}
\usepackage{lmodern}
\usepackage{fixltx2e}
\usepackage{microtype}
\usepackage{xspace}
\usepackage{enumitem}

\newcommand{\figureend}{\rule{\textwidth}{0.5pt}}

\makeatletter
\newcommand\etc{etc\@ifnextchar.{}{.\@}\xspace}

\makeatother

\usepackage{graphicx}

\newcommand{\inlinegraphic}[2]{
  \dimendef\grafheight=255\dimendef\grafvshift=254
  \grafheight=#1
  \grafvshift=-0.5\grafheight
  \advance\grafvshift by 0.5ex
  \raisebox{\grafvshift}{\includegraphics[height=\grafheight]{images/#2}\xspace}
}

\newcommand{\ninlinegraphic}[2][1.0]{
  \dimendef\grafheight=255\dimendef\grafvshift=254
  \setbox0 = \hbox{\scalebox{#1}{\includegraphics{images/#2}}}
  \grafheight=\the\ht0
  \grafvshift=-0.5\grafheight
  \advance\grafvshift by 0.5ex
  \raisebox{\grafvshift}{\includegraphics[height=\grafheight]{images/#2}\xspace}
}

\usepackage{latexsym}
\usepackage{amssymb}
\expandafter\let\csname equation*\endcsname\relax
\expandafter\let\csname endequation*\endcsname\relax
\usepackage{amsmath} 
\usepackage{stmaryrd}
\usepackage{mathtools}
\usepackage{stackrel}

\usepackage{amsthm}
\newtheorem{theorem}{Theorem}[section]

\theoremstyle{definition}\newtheorem{example}[theorem]{Example}
\theoremstyle{definition}
\theoremstyle{definition}\newtheorem{definition}[theorem]{Definition}
\theoremstyle{definition}
\theoremstyle{definition}
\theoremstyle{definition}

\newcommand{\sizeof}[1]{
  \left|#1\right|}

\newcommand{\ket}[1]{
    \ensuremath{\left|  #1 \right\rangle}\xspace}

\newcommand{\CZ}{CZ\xspace}
\newcommand{\CX}{CX\xspace}

\newcommand{\CNOT}{\CX}

\ifx\numericids\undefined

\else

\fi

\newcommand{\zxcalculus}{\textsc{zx}-calculus\xspace}
\newcommand{\zxdiagram}{\textsc{zx}-diagram\xspace}
\newcommand{\zxdiagrams}{\textsc{zx}-diagrams\xspace}

\usepackage{dsfont}

\newcommand{\tket}{\ensuremath{\mathsf{t}|\mathsf{ket}\rangle}\xspace}

\usepackage{subcaption}
\usepackage{csvsimple}
\usepackage{multirow}
\usepackage{makecell}
\usepackage{texlogos} 
\usepackage{algpseudocode}
\usepackage{ragged2e}

\DeclareFixedFont{\ttb}{T1}{txtt}{bx}{n}{10} 
\DeclareFixedFont{\ttm}{T1}{txtt}{m}{n}{10}  

\usepackage{color}
\definecolor{deepblue}{rgb}{0,0,0.5}
\definecolor{deepred}{rgb}{0.6,0,0}
\definecolor{deepgreen}{rgb}{0,0.5,0}

\usepackage{listings}
\newcommand\pythonstyle{\lstset{
language=Python,
basicstyle=\ttm,
otherkeywords={self},             
keywordstyle=\ttb\color{deepblue},
emph={MyClass,__init__},          
emphstyle=\ttb\color{deepred},    
stringstyle=\color{deepgreen},
frame=tb,                         
numbers=left,
numberstyle=\tiny\color{gray},
showstringspaces=false            
xleftmargin=10pt,
xrightmargin=10pt,
}}

\lstnewenvironment{python}[1][]
{
\pythonstyle
\lstset{#1}
}
{}

\newcommand\pythonexternal[2][]{{
\pythonstyle
\lstinputlisting[#1]{#2}}}

\newcommand\pythoninline[1]{{\pythonstyle\lstinline!#1!}}

\usepackage{hyperref}

\usepackage{tikzit}  

\newcommand{\inlinetikzfig}[2][1.0]{
  \dimendef\grafheight=255\dimendef\grafvshift=254
  \setbox0 = \hbox{\scalebox{#1}{\input{#2.tikz}}}
  \grafheight=\the\ht0
  \grafvshift=-0.5\grafheight
  \advance\grafvshift by 0.5ex
  \raisebox{\grafvshift}{\input{#2.tikz}}
}

\newcommand{\inltf}[1]{\InputIfFileExists{#1.tikz}{}{\input{./figures/#1.tikz}}}

\pgfdeclarelayer{background} 
\pgfdeclarelayer{foreground}
\pgfdeclarelayer{edgelayer}
\pgfdeclarelayer{nodelayer}
\pgfsetlayers{background,edgelayer,nodelayer,main,foreground} 

\tikzstyle{halfsize}=[x=0.5cm, y=0.5cm]
\tikzstyle{normalsize}=[]
\tikzstyle{doublesize}=[]

\tikzstyle{(null)}=[]
\tikzstyle{plain}=[]

\usepackage{tikzzx}
\usepackage{tikzcircuits}
\usepackage[undirected]{tikzquanto} 

\tikzstyle{double arrow}=[<->, >=triangle 45, line width=1pt]
\tikzstyle{single arrow}=[->, >=triangle 45, line width=1pt]

\usetikzlibrary{calc}

\tikzset{every picture/.style={line width=0.75pt}} 
\tikzset{directed edge/.style={postaction={decorate,decoration={markings,mark=at position 0.5 with {\arrow{>}}}}}}

\usepackage{rwd-drafting}
\usepackage{tikz}
\usepackage{csvsimple}
\usepackage{booktabs}

\begin{document}


\title{\tket : A Retargetable Compiler for NISQ Devices}

\author{Seyon Sivarajah$^1$$^{*}$\footnote[1]{seyon.sivarajah@cambridgequantum.com},
  Silas Dilkes$^1$$^{*}$, 
  Alexander Cowtan$^1$$^{*}$,\\ 
  Will Simmons$^{1}$$^{*}$,
  Alec Edgington$^1$,
  and Ross Duncan$^{1,2}$}

\address{$^1$ Cambridge Quantum Computing Ltd, 9a~Bridge Street,
  Cambridge, United Kingdom}
\address{$^2$ Department of Computer and Information Sciences,
  University of Strathclyde, \\\ \ \ 26~Richmond Street, Glasgow, United
  Kingdom}
\vspace{3mm}
\address{$^{*}$ These authors contributed equally to this work.}


\begin{abstract}
We present \tket, a quantum software development platform produced by Cambridge
Quantum Computing Ltd.  The heart of \tket is a language-agnostic optimising
compiler designed to generate code for a variety of NISQ devices, which has
several features designed to minimise the influence of device error.  The
compiler has been extensively benchmarked and outperforms most competitors in
terms of circuit optimisation and qubit routing.
\end{abstract}

%
%
%


\section{Introduction}
\label{sec:intro}

Quantum computing devices promise significant advantages for a wide variety of
information processing
tasks~\cite{Shor:PolyTimeFact:1997,Grover:1997qc,PhysRevLett.103.150502}. For
some tasks, notably the simulation of condensed-matter physics, the abstract
structure of the problem may be sufficiently similar to the physical structure
of the device that translation from one to the other is natural and (relatively)
straightforward~\cite{Georgescu:2013aa}. However, for most problems, and most
quantum computers, this is not the case.  Quantum algorithms are often described
in terms that facilitate proving correctness or deriving asymptotic complexity
estimates, without reference to a specific computing device on which to execute
them.  The translation from a high-level description of the algorithm to a
machine-specific sequence of physical operations is called \emph{compilation},
and is essential to realising the supposed computational advantage of quantum
algorithms.

In computer science, the term \emph{compiler} was introduced by Grace Hopper in
the early 1950s~\cite{10.1145/609784.609818}, and originally referred to a
routine which ``compiled'' a desired program from pre-existing pieces.  Today 
the term
denotes a program that translates a human-readable programming language into the
binary language of the machine that will execute it.  Early compilers produced
code that was grossly inefficient, compared to what an average human programmer
could write; however, today's sophisticated \emph{optimising} compilers reliably
generate code that runs more quickly and uses less memory than even the best
human programmer could manage~\cite{Godbolt2019}.  The steady improvement of
compiler technology has, in turn, enabled programming languages to increase in
power and sophistication, increasing the conceptual distance between the
programmer and the executable machine language.

By comparison, almost all programming systems available for quantum computing
are conceptually primitive, remaining extremely close to the basic quantum
circuit model~\cite{Knill:Pseudocode:1996}.  Although higher-level
application-oriented toolkits are becoming
available~\cite{ibm:qiskit-aqua,Bergholm:aa}, the programmer must usually
describe the algorithm to be run in terms of basic unitary gates.  On the other
hand, quantum computing hardware displays great diversity.  Superconducting and
ion-trap-based quantum processors are now available from multiple commercial
companies~\cite{IBMQuantumExp,rigetti-systems,google-systems,Wright:2019aa,honeywellquantumsystems},
while other technologies such as photonics are not far
behind~\cite{xanadu-hardware,Rudolph:2016aa}.  Different underlying technologies
have very different performance parameters and trade-offs, and even broadly
similar devices may differ in what basic operations are available.  Even in the
context of the simple, circuit-centric, programming model, the requirement to
translate an abstract circuit into something suitable for the chosen device
creates the need for a compiler.  Naive approaches to this translation can
significantly increase the size of the circuit; therefore the other major task
for a quantum compiler is \emph{circuit optimisation}, to minimise the resources
required by the program.

Circuit optimisation is especially pertinent on so-called \emph{noisy
intermediate-scale quantum} (NISQ) devices.  Preskill
\cite{Preskill2018quantumcomputingin} defines a NISQ device as having a memory
size of 50--100 qubits, and sufficient gate fidelity to carry out around 1000
two-qubit operations with tolerable error rates.  We will adopt a wider
definition: a NISQ device is any quantum computer for which general-purpose
quantum error correction~\cite{NieChu:QuantComp:2000} is not feasible and
hardware errors are expected.  Because of these ineradicable errors, mere qubit
count is poor measure of the capability of NISQ devices. The longer the
computation runs, the more noise builds up. 

NISQ devices therefore impose strict limitations both on the number of qubits
available to algorithms and on the maximum circuit depth that can be achieved. 
Aside from the obvious requirement to use this limited hardware budget in the
most efficient manner possible, the noisiness of NISQ machines introduces
further complications.  Since many common textbook algorithms such as quantum
phase estimation \footnote{But see O'Brien \etal~\cite{O_Brien_2019}.} are not
feasible in the available circuit depth, hybrid algorithms such as the
Variational Quantum Eigensolver (VQE)~\cite{Peruzzo:2014aa} and the Quantum
Approximate Optimisation Algorithm (QAOA)~\cite{Farhi:2014aa} have been proposed
instead.  While the circuit depths required by these algorithms are more
favourable to NISQ devices, they are based on repeatedly executing circuits
inside a classical optimisation loop, where both the rate of convergence and the
accuracy of the final result can be adversely affected by device noise.  In
consequence, any compiler for NISQ devices should aim to maximise the overall
fidelity of the computation.  Minimising the number of operations helps, but
other techniques may be
employed~\cite{Kandala:2018ab,PhysRevA.94.052325,PhysRevLett.121.220502,Ball:2020aa}.

This paper describes \tket, a compiler system for NISQ devices that aims to
achieve these objectives.  The core of \tket is a flexible optimising compiler
which supports multiple programming frameworks, and multiple quantum devices. 
It is specifically designed for NISQ devices, and includes features that
minimise the influence of device errors on computation.  As we demonstrate in
Section~\ref{sec:benchmarks}, \tket's optimisation and qubit mapping routines
reliably outperform other compilers.  The system also includes runtime
management features to facilitate the variational algorithms typical of NISQ
devices.

\subsection{NISQ devices and their software}

Before addressing the \tket system, we consider a schematic variational
algorithm in the context of the system architecture of an idealised NISQ device.
For purposes of illustration, a toy VQE algorithm is shown in
Figure~\ref{fig:vqe-pseudocde}.

\begin{figure}[t]
  \begin{algorithmic}
    \State $C(\cdot) \gets $ GenerateParameterisedCircuit()
    \State $M \gets $ GenerateListOfPauliMeasurements() 
    \State $\overline\theta \gets \overline\theta_0$ 
    \State $E \gets 0$
    \While {$E$ has not converged} 
    \For {$i$ from 1 to $N\_ITERS$} 
    \For {$m \in M$} 
    \State $r_{m,i} \gets $ Execute$(C(\overline\theta);m)$
    \EndFor
    \EndFor
    \State $E' \gets $ EstimatorFunction$(\overline r)$ 
    \State $\overline\theta \gets $ ClassicalOptimiser$(\overline\theta,E,E')$ 
    \State $E \gets E'$
    \EndWhile
    \State \Return $E$
  \end{algorithmic}
  \caption{The typical structure of a variational quantum algorithm}
  \label{fig:vqe-pseudocde}
\end{figure}

The first point to note in this example is that the central
``Execute$(C(\overline\theta);m)$'' subroutine is the only part that runs on the
quantum device: the other subroutines and the main loop are classical.  The
subroutines ``EstimatorFunction'' and ``ClassicalOptimiser'' are used repeatedly
inside the main loop -- the characteristic of a hybrid algorithm -- and their
outputs are used in the next quantum execution.  The first two subroutines,
``GenerateParameterisedCircuit'' and ``GenerateListOfPauliMeasurements'', are
tasks that are usually considered part of the compiler, but observe that inside
the main loop a fresh quantum circuit must be built using the parameterised
circuit $C(\cdot)$, the measurement $m$, and the new parameters
$\overline\theta$.  How does an algorithm like this map onto a realistic quantum
computer system?

\begin{figure}
  \begin{center}
  \includegraphics[scale=0.8]{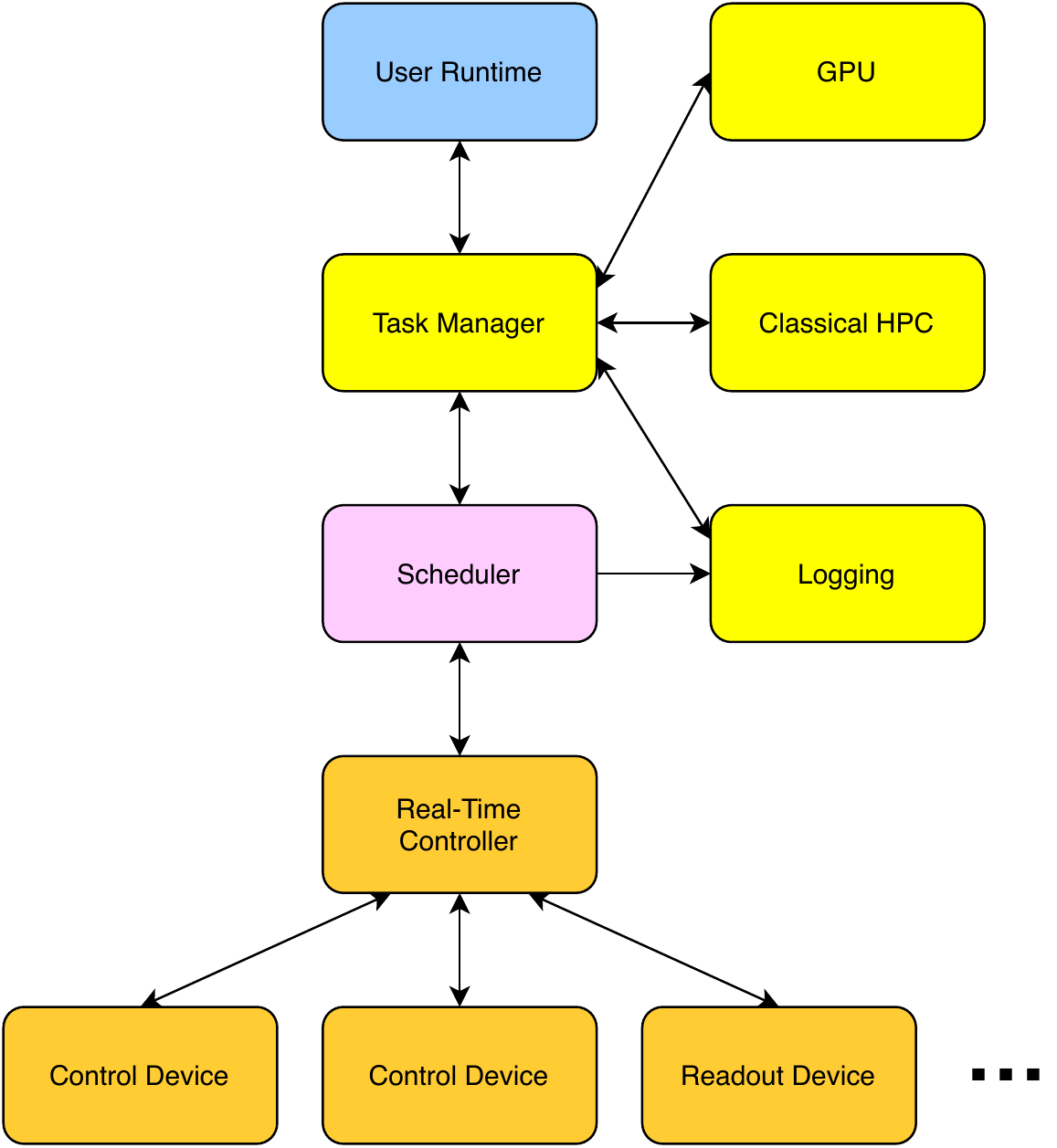}
  \end{center}
  \caption{Idealised system architecture for a NISQ Computer}
  \label{fig:NISQsysarch}
\end{figure}

While it is common to talk about a ``quantum computer'' as a unified device, in
practice it consists of multiple subsystems, each of which is a computer in its
own right.  Running a quantum algorithm therefore involves a large number of
software components in a mixture of runtime environments, with very different
performance demands.  Figure~\ref{fig:NISQsysarch} displays a realistic
architecture for a NISQ computer.  The lowest level comprises the programmable
\emph{devices} which drive the evolution of the qubits and read out their
states.  An example of this kind of device is an arbitrary waveform generator,
as found in many superconducting architectures.  The microwave pulse sequences
output by these devices are generated by simple low-level programs optimised for
speed of execution.  These devices, and the \emph{real-time controller} which
synchronises them, operate in a hard real-time environment where the computation
takes place on the time-scale of the coherence time of the qubits.  These
components combine to execute a single instance of a quantum circuit, possibly
with some classical control. By analogy with GPU computing, we refer to this
layer as a \emph{kernel}.

One level higher, the \emph{scheduler} is responsible for dispatching circuits
to be run and packaging the results for the higher layers. It is also likely to
be heavily involved in the device calibration process. (Calibration data are an
important input to the compiler.)  This layer and those below may be thought of
as the low-level system software of the quantum computer, and must normally be
physically close to the device.  In the layer above we find service-oriented
middleware, principally the \emph{task manager}, which may distribute jobs to
different quantum devices or simulators, GPUs, and perhaps conventional HPC
resources, to perform the various subroutines of the quantum algorithm.  This
layer may also allocate access to the quantum system among multiple users. 
Finally, at the highest level is the \emph{user runtime}, which defines the
overall algorithm and integrates the results of the subcomputations to produce
the final answer.

With this picture in mind, we see that the path from a high-level program
describing a quantum algorithm to its final result involves many stages of
decomposition and compilation in order to run in this heterogeneous environment.
In practice, some of these stages may be amalgamated or absent.  In this paper
we will focus on the generation of the kernel, since this is the indispensable
part of the process, and can be (to some extent) decoupled both from the
high-level architecture and the low-level system-specific parts.

The picture is further complicated when considering quantum computers capable of
error correction.  The ``logical'' kernel must be translated to an encoded
equivalent; subroutines to perform gate synthesis must be added; and error
detection correction stages must be interleaved in the main algorithm.  However,
every part of the NISQ process is also required in the error-corrected case, so
we focus on compilation in the NISQ context.

\subsection{Related work}
\label{sec:related-work}

The last few years have seen an explosion of interest in quantum programming
languages, and the problems of quantum compilation have been explored at various
levels of abstraction~\cite{Hner2016A-Software-Meth}, from high-level algorithm
design to pulse control at the machine level.

Several languages have been developed for quantum programming.
Quipper~\cite{Alexander-S.-Green:2013fk} is a functional language for quantum
circuits embedded in Haskell. The ScaffCC compiler~\cite{JavadiAbhari20152},
based on the LLVM framework, compiles an extension of C, and can be configured
for routing to specific architectures. Q\#~\cite{Svore:2018:QES:3183895.3183901}
is a hybrid classical--quantum language designed to facilitate the development
of programs that can be run on a simulator (and eventually on actual hardware).
Strawberry Fields~\cite{Killoran2019strawberryfields} is a Python-based quantum
programming framework that is based on the ``continuous variable'' (CV) model of
computation.

Several other compilation systems have been developed as Python modules
targeting specific hardware. These include
the Forest SDK/pyQuil~\cite{forest} (for Rigetti backends),
Qiskit~\cite{ibm:qiskit} and ProjectQ~\cite{Steiger2016ProjectQ:-An-Op} (for IBM
backends), and Cirq~\cite{cirq} (for Google backends). Other projects have
adopted a backend-agnostic approach. XACC~\cite{XACC} is a quantum programming
framework that can target several different backends as plug-ins.
TriQ~\cite{Murali:2019aa} uses ScaffCC to compile quantum software for several
different architectures in order to study their performance characteristics. 
Even for the full-stack systems, the compiler element (e.g. Quilc~
\cite{robert_s_smith_2020_3677537} from the Forest SDK or the transpiler passes 
in Qiskit Terra) can be invoked to compile for arbitrary devices.

A range of gate-level circuit-optimisation techniques have been explored,
including the use of phase polynomials~\cite{Nam:2018aa} and constraint
programming~\cite{Venturelli2019QuantumCC}. There are also promising results for using information on the noise characteristics and fidelities of the target device to assist compilation~\cite{Murali:2019ab,murali2020software,peterson2019fixeddepth}. Meanwhile at the level of machine
control there have been efforts to optimise the implementation of variational
algorithms using automatic differentiation and interleaving compilation with
execution~\cite{PhysRevA.95.042318,Gokhale:2019:PCV:3352460.3358313}.

\subsection{Synopsis}
\label{synopsis}

Here we give an overview of the \tket system. Subsequent sections give detailed
descriptions of the modular front- and back-ends, the intermediate
representation used, the transform system, some of the optimisation methods, and
the system of qubit placement and routing. Section~\ref{sec:benchmarks} provides
comparative benchmark results for the performance of the optimiser and qubit
allocation engine.  The benchmark set of circuits is available for download.

\subsection{How to get \tket}
\label{sec:how-get-tket}

\begin{figure}[t]
  \begin{center}
  \includegraphics[width=\textwidth]{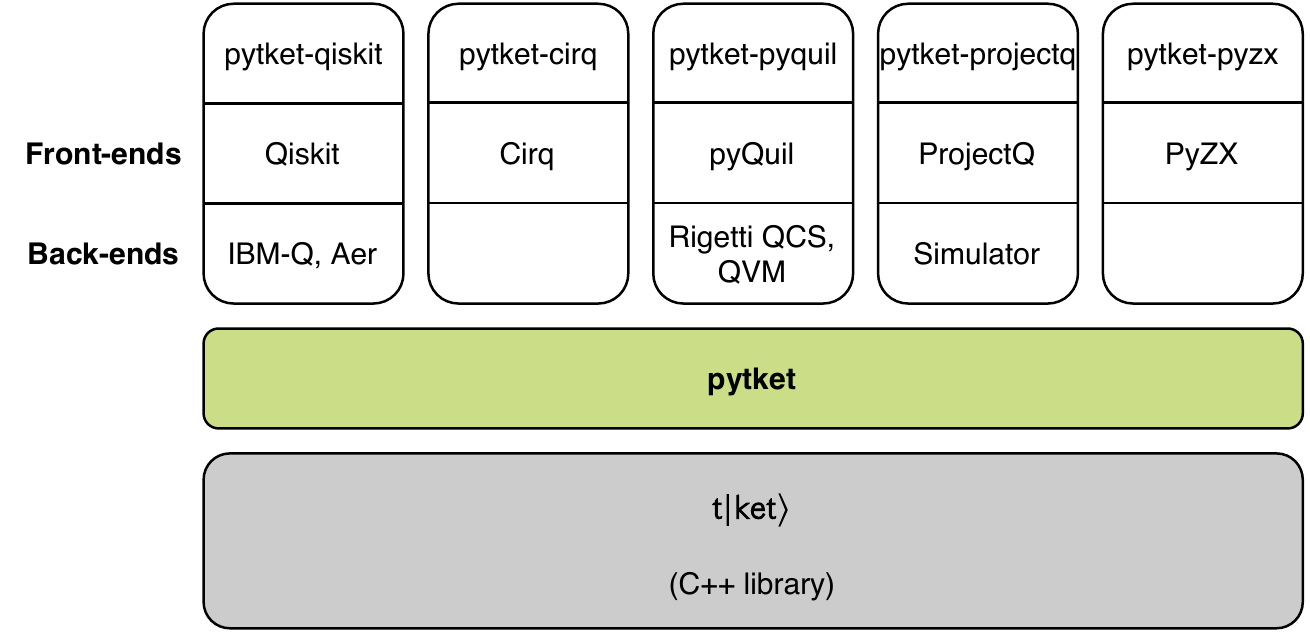}
  \end{center}
  \caption{Components of \tket}
  \label{fig:install}
\end{figure}

While the core of \tket is a highly optimised \cpluspluslogo\ library, the
system is available as the Python module \texttt{pytket}, which provides the
programming interface and interoperability with other systems.  It can be
installed on Linux and MacOS using the command:
\begin{lstlisting}
  pip install pytket
\end{lstlisting}
and the documentation is available online at:
\url{https://cqcl.github.io/pytket/}. Figure~\ref{fig:install} shows the
components that are installed by this command.

To interface with other software packages, and to use back-ends that depend on
external software, the user must also install plug-in packages. At the time of
writing, the available plug-ins are: \texttt{pytket\_qiskit},
\texttt{pytket\_cirq}, \texttt{pytket\_pyquil}, \texttt{pytket\_projectq}, and
\texttt{pytket\_pyzx}. All of these can be installed using \texttt{pip} in the
same manner as the core package. The \texttt{pytket} module is free for
non-commercial use. We encourage the reader to try it out for themselves!

\section{System Overview}
\label{sec:overview}

The \tket system consists of two main components: a powerful optimising compiler
written in \cpluspluslogo, and a lightweight user interface and runtime system
written in Python.  This Python layer allows the user to define circuits and
invoke compiler functions, while the runtime environment marshalls and
dispatches kernels for execution, and provides convenience methods for defining
variational loops, updating parameters, and collating statistics across circuit
evaluations.  Optional Python extensions provide interfaces to third-party
quantum software systems.  The overall structure is illustrated in
Figure~\ref{fig:install}.

In the classical setting, a compiler translates a human-readable programming
language into machine-executable object code.  This process can be divided into
three stages: a \emph{front-end}, which handles lexing, parsing, semantic
analysis, and other tasks which depend on the source language; a
\emph{back-end}, which allocates registers and generates suitable instruction
sequences in the target machine language; and an intermediate stage, which
performs data and control-flow analysis on an \emph{intermediate representation}
(IR) of the program, which is independent of both the source and the target
languages.  Modern compiler systems, such as LLVM~\cite{llvm:website}, use a
standard IR to decouple these three stages, making it relatively simple to add
support for a new programming language or machine architecture to an existing
compiler framework.

\begin{figure}[h]
  \begin{center}
  \includegraphics[scale=0.6]{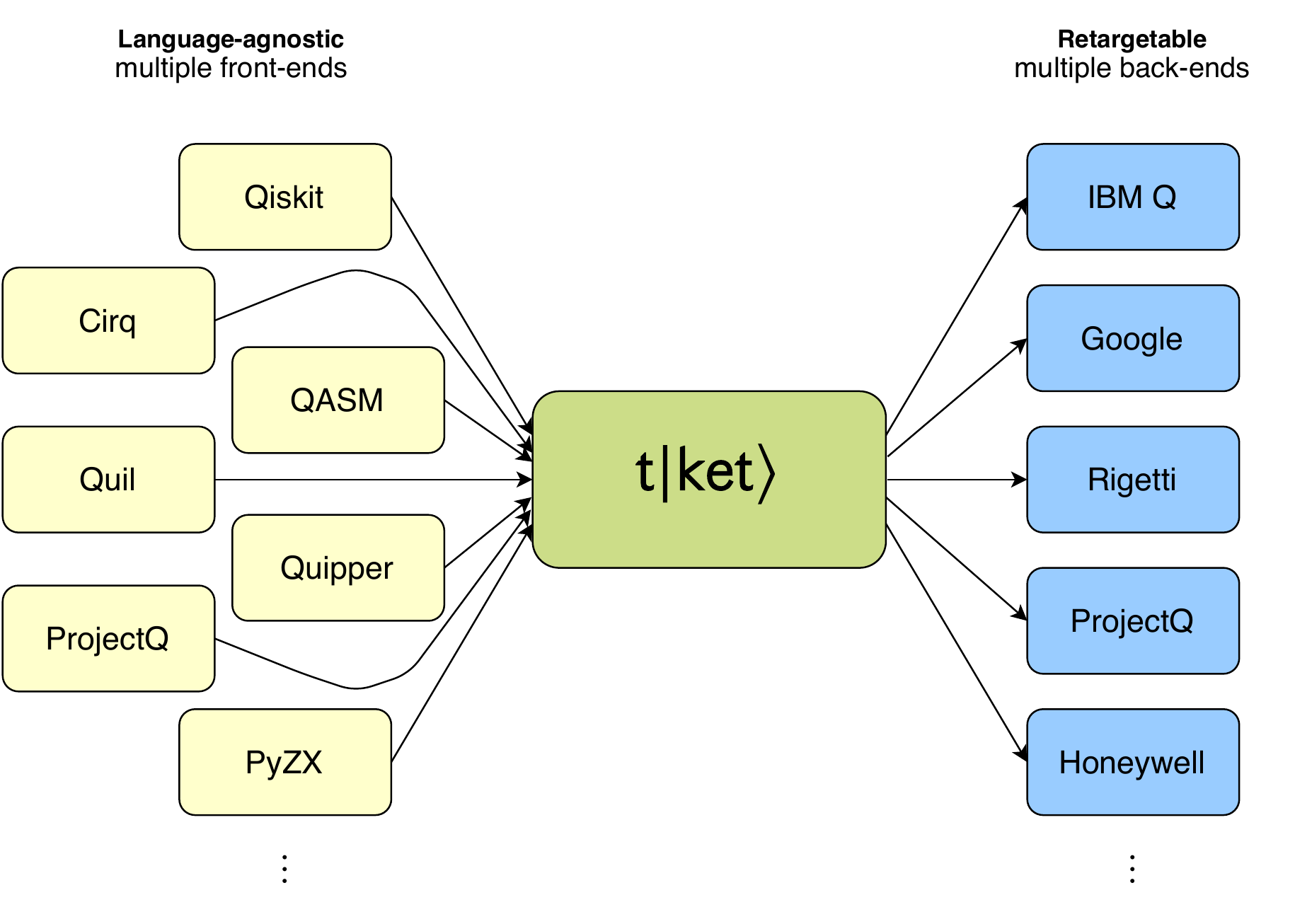}
  \end{center}
  \caption{Modular front-ends and back-ends for \tket}
  \label{fig:inout}
\end{figure}

\tket was designed from the ground up to be \emph{retargetable}, meaning that it
can generate code for many different quantum devices, and \emph{language
agnostic}, meaning that it accepts input from most of the major quantum software
platforms.  For this reason, its overall structure, shown in
Figure~\ref{fig:inout}, follows the same basic pattern as the LLVM.  A variety
of lightweight front-end units translate the desired input language into the
\tket IR.  This internal representation is a generalisation of the usual
language of quantum circuits based on hierarchical non-planar maps; this is
described in more detail in Section~\ref{sec:circuits}.  Standard quantum
circuits are easily embedded into the \tket IR -- a fact which eases the task of
adding new front-ends -- but many node types that are not unitary gates are also
available.

Once the input has been translated to the IR, the central circuit transformation
engine can begin its work.  The transformation engine performs a
user-configurable sequence of rewrites of the IR; some examples are described in
Section~\ref{sec:circ-optim-meth}. Typically this proceeds in two phases: an
architecture-independent optimisation phase, which aims to reduce the size and
complexity of the circuit; and an architecture-dependent phase, which prepares
the circuit for execution on the target machine.  This phase itself decomposes
into a \emph{rebase}, which maps the gates present in the circuit into those
supported by the device, and a \emph{qubit mapping} phase.  The mapping phase is
necessary to ensure that all qubits that are required to interact during the
program are physically able to do so; this typically increases the size of the
circuit, since most devices do have restricted interactions between their
qubits. This is described in detail in Section~\ref{sec:routing}.  

The end product of this process is a \emph{kernel}: a circuit that can be
executed on the chosen target device.  The kernel may then be scheduled for
execution by the runtime environment, or simply saved for later.

In keeping with its focus on NISQ devices, the design of \tket is minimalistic
compared to the schema proposed by H{\"a}ner
\etal~\cite{Hner2016A-Software-Meth}.  There is no error correction, and \tket
does not include a linker, preferring to rely on application programming
frameworks such as CQC's Eumen or IBM's Qiskit Aqua to provide libraries of
common routines.  The lowest layer of compilation -- translation of kernels to
control signals for the lasers, microwave generators, and so on -- is left to
the hardware implementor. However, since \tket takes into account the device's
architectural constraints during the compilation phase, this last stage of
translation can be minimal.

Thanks to its retargetability, \tket can be used as a cross-compiler: source
programs produced from any supported front-end can be compiled to run on
hardware produced by any vendor.

\section{Front-ends and Back-ends}
\label{sec:front-ends-back}

The \texttt{pytket} interface can be used directly to build quantum
circuits from individual gates in the standard way.  While this may be
acceptable for small experiments, more powerful high-level tools are
preferable for larger or more complex tasks.  For this reason \tket
sports a range of lightweight front-end modules for different quantum
programming systems.  The industry-standard
OpenQASM~\cite{Cross2017Open-Quantum-As} and the functional language
Quipper~\cite{Alexander-S.-Green:2013fk} are supported via direct
source-file input. Python converters provide support for IBM's
Qiskit~\cite{ibm:qiskit}, Google's Cirq~\cite{cirq} and Rigetti's
pyQuil~\cite{forest}, as well as the independent open-source projects
ProjectQ~\cite{Steiger2016ProjectQ:-An-Op} and
PyZX~\cite{Kissinger:2019ab}. These Python libraries in turn support
higher-level application programming frameworks such as
OpenFermion~\cite{McClean:2017aa} or Rigetti Grove. Support for
Q\#~\cite{Svore:2018:QES:3183895.3183901} is planned for the next
release.

\begin{figure}[t]
    \pythonexternal[xleftmargin=.1\textwidth, xrightmargin=.1\textwidth]{code/backends.py}
    \caption{Code example showing front-end and back-end use. A circuit is read
    in from a QASM file; operations are appended to it using the \texttt{pytket}
    interface; the circuit is compiled to satisfy the constraints of a back-end,
    and then executed. The \texttt{IBMQBackend} class is included in the
    \texttt{pytket\_qiskit} extension package.}
    \label{code:backends}
    \figureend
\end{figure}

Similarly, \tket offers multiple back-ends, each supporting a different quantum
hardware platform or classical simulator. Supporting a given platform implies,
firstly, generating a circuit that respects the constraints of the hardware or
simulator (generally, connectivity and primitive gate limitations); secondly,
the back-end must dispatch the kernel for execution and collate results. The
first of these tasks is handled by the system described in
Section~\ref{sec:transform}. Each back-end class provides a default compiler
pass, which guarantees that a compiled circuit will respect the relevant
constraints.

\tket attempts to provide a uniform interface across the various back-end
platforms, so that a user can easily change back-ends for an experiment without
changing anything else in their code.  At the time of writing, \tket supports
all IBM~Q and Rigetti devices via their online access services, and experimental
devices produced by Honeywell Quantum Systems, Oxford Quantum Circuits and the
University of Maryland.  In addition, \tket can use the ProjectQ, IBM Aer and
Rigetti QVM simulators.  Various other machines are supported indirectly using
either QASM output or \tket's integration with Qiskit and Cirq. 
Figure~\ref{code:backends} shows an example of front-end input followed by
circuit compilation, execution and result retrieval via a back-end. For
back-ends that support it, circuit submission and job retrieval can be performed
separately, allowing asynchronous execution of the quantum circuit.

Utility functions are also provided for postprocessing of results, such as
calculation of expectation values. Generic mitigation of classical
state-preparation-and-measurement (SPAM)~\cite{spam} errors across back-ends is
scheduled for release in 2020.

\section{Representing Circuits}
\label{sec:circuits}
\begin{figure}[bt]
  \begin{center}
  \includegraphics[scale=0.5]{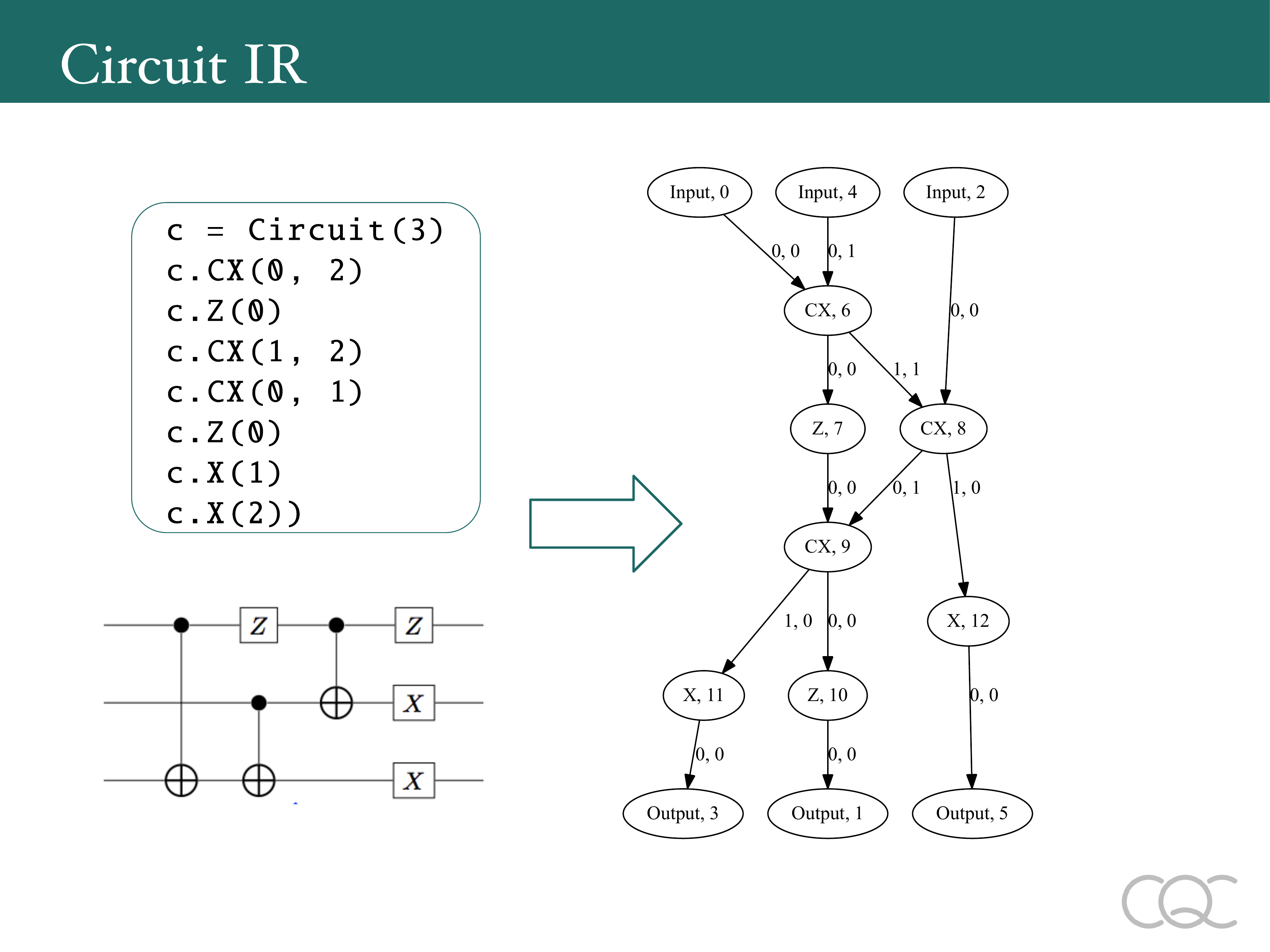}
  \end{center}
  \caption{\tket circuit internal representation. Pairs of values
  labeling edges correspond to port numbers at the source and target
  vertices. Note that the Input and Output label numbers are even and
  odd respectively, so that qubit 0 corresponds to the path from
  ``Input, 0'' to ``Output, 1'', qubit 1 is the path from ``Input, 2'' to
  ``Output, 3'', and so on.}
  \label{fig:port_graphviz_diagram}
  \figureend
\end{figure}

The standard intermediate representation (IR) in \tket is the
\emph{circuit}. A circuit is a labelled directed acyclic graph (DAG)
with some additional structure.  Vertices in the DAG correspond to
operations, usually quantum or classical logic gates, but also
\emph{boxes}, a kind of opaque container which we will define later,
and certain compiler-internal meta-operations.  Edges in the DAG track
the flow of computational resources from operation to operation.
Typically, these resources are qubits, and the operations are unitary
gates.  Since many operations do not act symmetrically on their
inputs, we add port labels for the incoming and outgoing edges at each
vertex to distinguish between, for example, the control and target
qubits of a \CNOT gate.  Each input port of a quantum operation is
paired with an output port, allowing the path of a particular resource
unit to be traced through the circuit, as shown in
Figure~\ref{fig:port_graphviz_diagram}. 

\begin{figure}[tb]
  \begin{center}
  \includegraphics[scale=0.5]{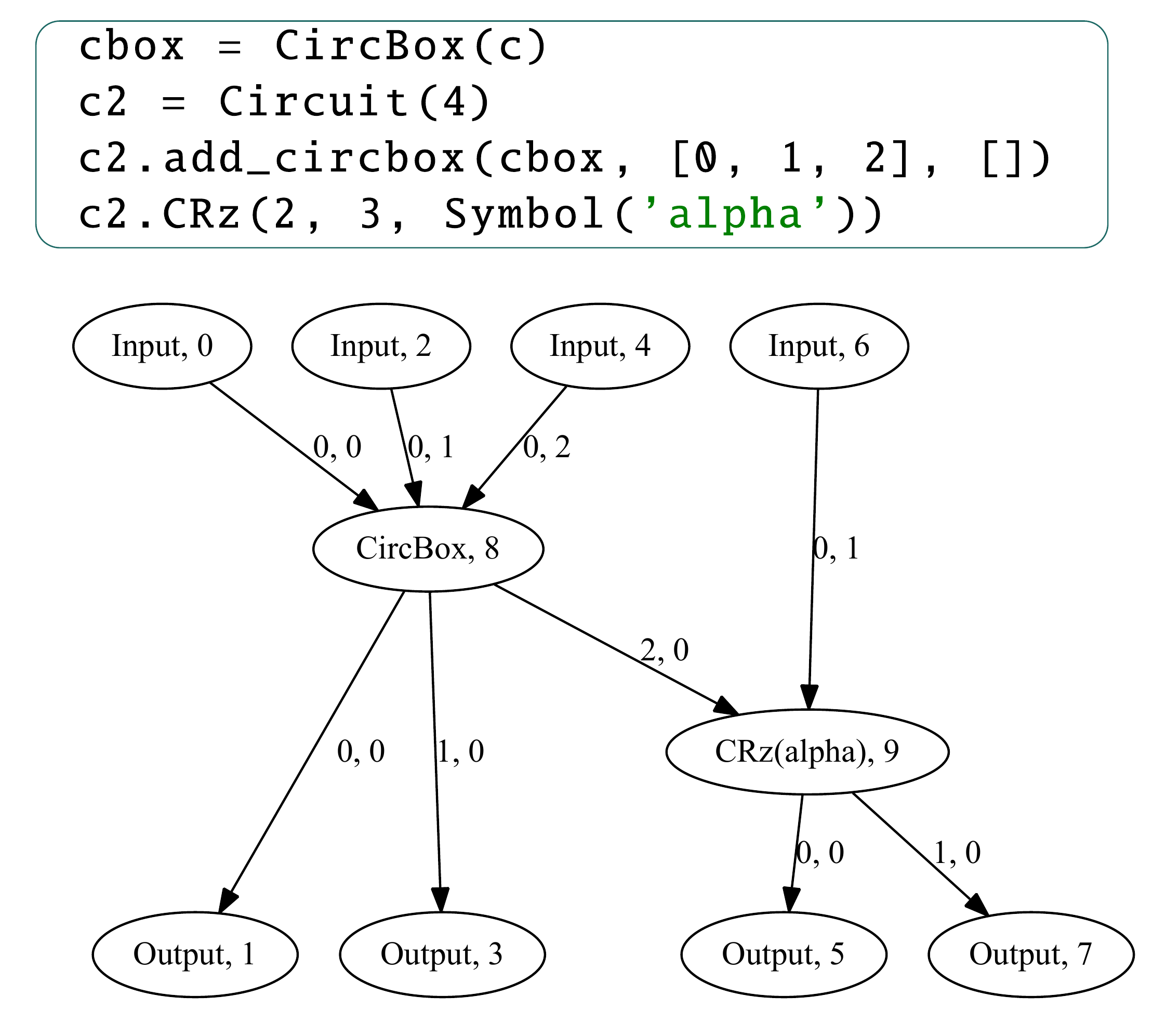}
  \end{center}
  \caption{\tket circuit with a box containing the circuit from
  Figure~\ref{fig:port_graphviz_diagram}. A parameterised controlled-Rz gate
  is also shown, with a symbolic parameter.}
  \label{fig:box_graphviz_diagram}
  \figureend
\end{figure}

For qubits, this linear resource management is justified by the
no-cloning~\cite{Wootters1982A-single-quantu} and
no-deleting~\cite{Pati2000Impossibility-o} theorems.  However, for classical
operations, this effectively means that output values must overwrite their
previous value. This treatment of classical bits is an artefact of the
simplified classical computational model of QASM~\cite{Cross2017Open-Quantum-As}
and devices that have adopted it, requiring explicit allocation of classical
registers (both for classical input to execution and for result retrieval) and
forbidding dynamic allocation of scratch space.

At the input and output boundaries of a circuit, the resource units --
which we may identify with storage locations -- are partitioned into
\emph{registers}. A circuit can contain arbitrarily many registers,
and resource units are represented within a register by identifiers that
are unique within the circuit. The registers specify an ordering of
the resource units \footnote{In category-theoretic terms, the triple
  (I,O,G) of an input ordering, output ordering and graph
  corresponds to a structured cospan~\cite{baez:2019aa}, where G is
  the apex.}. This ordering allows circuits to be composed
sequentially and in parallel, and act as ``port labels'' for entire
circuits, just as individual operations have ports within a
circuit. This means that the process of composing circuits is
identical to the composition of individual operations.

\subsection{Gate Types}
\begin{table}[h!]
\centering
\begin{tabular}{|c|c|} 
 \hline
 Class of operation & Example \\
 \hline
 Basic single-qubit gate & Hadamard  \\ 
 Parameterized single-qubit gate & \(\mathrm{Rz}(\alpha)\)  \\
 Basic two-qubit gate & \CNOT  \\
 Parameterized two-qubit gate & \(\mathrm{CRz}(\alpha)\) \\
 Basic multi-qubit gate & \(\mathrm{C_nX}\) \\
 Parameterized multi-qubit gate & \(\mathrm{C_nRy}(\alpha) \)\\ 
 Classical output gate & Measure \\
 Meta-operation & Barrier \\
 Hierarchical vertex & CircBox \\
 [1ex] 
 \hline
\end{tabular}
\caption{Classes of operations available for circuits in \tket}
\label{table:gate_desc}
\end{table}

There are a wide array of allowed logic gates in \tket, covering the native
gates of the platforms that \tket can interface with. An overview of the kinds
of supported gates is given in Table~\ref{table:gate_desc}. The most common
quantum gates are one- and two-qubit gates, reflecting the native gates on
physical superconducting and ion-trap hardware, but some gates with arbitrary
quantum controls are allowed; these must eventually be decomposed to
hardware-native gates by the transform engine. All quantum gates in \tket can
have arbitrary classical control, and primitive classical logic gates are
supported. However, adding classical control to gates can limit the ability of
the rewrite engine to optimise the circuit. An enumeration of all the allowed
operation types in \tket can be found in the documentation at
\url{https://cqcl.github.io/pytket/build/html/optype.html}.

\emph{Boxes} are a special class of operations in \tket. A box vertex is a
container which encapsulates a whole circuit. In
Figure~\ref{fig:box_graphviz_diagram}, the circuit from
Figure~\ref{fig:port_graphviz_diagram} is put into a box within another circuit.
 Boxes allow for front-ends to take in high-level descriptions with subroutines.
 As this subcircuit can also contain box vertices, a single circuit can contain
a hierarchy of arbitrary rank. The hierarchy must be decomposed at the kernel
generation stage, but this decomposition is trivial because of the compositional
structure whereby circuits are equivalent to individual operations. As well as
explicit circuits, box vertices can also contain other representations, which
can be useful for optimising certain classes of quantum circuits. Because boxes
are opaque, the parent circuit is undisturbed by the optimisation procedure
acting on the subroutine. The next release of \tket will include such
optimisations acting directly on boxes.

\subsection{Gate Parameters}
Unlike large-scale, fault-tolerant quantum computers, NISQ devices generally
allow arbitrary angles on parameterised gates. Accordingly, \tket allows
arbitrary angles on all parameterised gates, up to IEEE~754
double-precision~\cite{IEEE:2019aa}. 

In Section~\ref{sec:intro} we briefly described the variational hybrid
quantum--classical algorithms proposed for NISQ devices. To enable the efficient
compilation of this class of algorithms, \tket supports \emph{symbolic}
parameters. This allows the compilation of a parameterised circuit corresponding
to an entire variational algorithm without requiring repeated compilation from
scratch at each iteration of the classical optimiser. The circuit in
Figure~\ref{fig:box_graphviz_diagram} contains a parameterised controlled-Rz
gate with a symbolic parameter. This class of circuits is handled using
\emph{partial compilation}: the circuit is precompiled with unknown, symbolic
parameters using an expressive symbolic manipulation library. The result can be
used as a template circuit and, after parameter values at a given iteration have
been substituted, further simple circuit rewriting can be performed before the
resulting kernel is sent to a backend to be run. This minimises the computation
required between iterations of the classical optimiser, reducing the overall
runtime of a variational algorithm while still using the rewrite engine of \tket
to minimise the resource costs of the circuits. The implementation of the
circuit class uses local adjacency lists at each vertex to allow near
constant-time edge and vertex insertion and removal.

\section{The \tket Transform System}
\label{sec:transform}

In general, a quantum algorithm can be expressed in multiple ways using a given
gate set; the goal is to express it in a way that minimises the gate count and
circuit depth. The field of circuit optimisation is well developed, with a
variety of optimisation strategies employed for different algorithms and target
hardware devices~\cite{Nam:2018aa,maslov2016basic,PhysRevA.88.052307}. Most
commonly, a circuit can be rewritten using unitary equality between circuits,
where a resource-inefficient subcircuit can be found and replaced using a
closer-to-optimal one \footnote{The process of finding and replacing subgraphs
in this manner is called \emph{double pushout rewriting}. See Ehrig
\etal~\cite{Ehrig:2006ab}.}. The core of \tket is a high-performance circuit
rewriting engine, referred to as the \emph{transform system}. A function that
performs rewrites using this system is called a \emph{transform pass}. Circuit
optimisation in \tket is described in more detail in
Section~\ref{sec:circ-optim-meth}. Aside from optimisation, the transform system
has an essential role in generating circuits that are executable on the target
hardware.

Each backend that \tket can target has associated with it a series of properties
that any valid circuit must satisfy. This will include, as a minimum, the set of
supported gates; for many architectures it will also include a graph
representing the connectivity between the qubits. Mapping logical qubits to
physical qubits also requires rewriting the circuit so that the interactions
between qubits correspond only to edges in the associated connectivity graph.
The transform pass that performs this rewriting is described in
Section~\ref{sec:routing}. Other properties may also be required, depending on
the platform, and these also require transform passes. 

Transform passes are composed sequentially; the resulting function is also a
transform pass. For instance, a typical compiler flow will consist first of some
optimisation on the circuit that has no regard for connectivity graph or gate
set, followed by passes that bring the circuit closer to satisfying all of the
constraints. Only if a circuit satisfies all these properties can it be executed
on the target hardware.

To document and constrain the composition of transforms, the \tket transform
engine implements a simple expression language, which follows the same
principles as ``contracts'' in object-oriented programming~\cite{contract_2001}.

The functions that verify that properties are satisfied are called
\emph{predicates}.  Each predicate is a function from a circuit to a boolean
value: \emph{true} if the circuit satisfies the corresponding property and
\emph{false} otherwise.  These functions can incorporate some external
information about the target hardware, such as connectivity graph and desired
gate set; when external information is required the predicates are generated by
higher-order functions. For example, to verify that the connectivity graph of a
specific architecture is satisfied, a higher-order function will take in a
connectivity graph and return the corresponding predicate. The full list of
allowed predicates is documented at
\url{https://cqcl.github.io/pytket/build/html/predicates.html}.

\begin{figure}[ht]
    \begin{subfigure}{\textwidth}
  \begin{center}
  \includegraphics[scale=0.3]{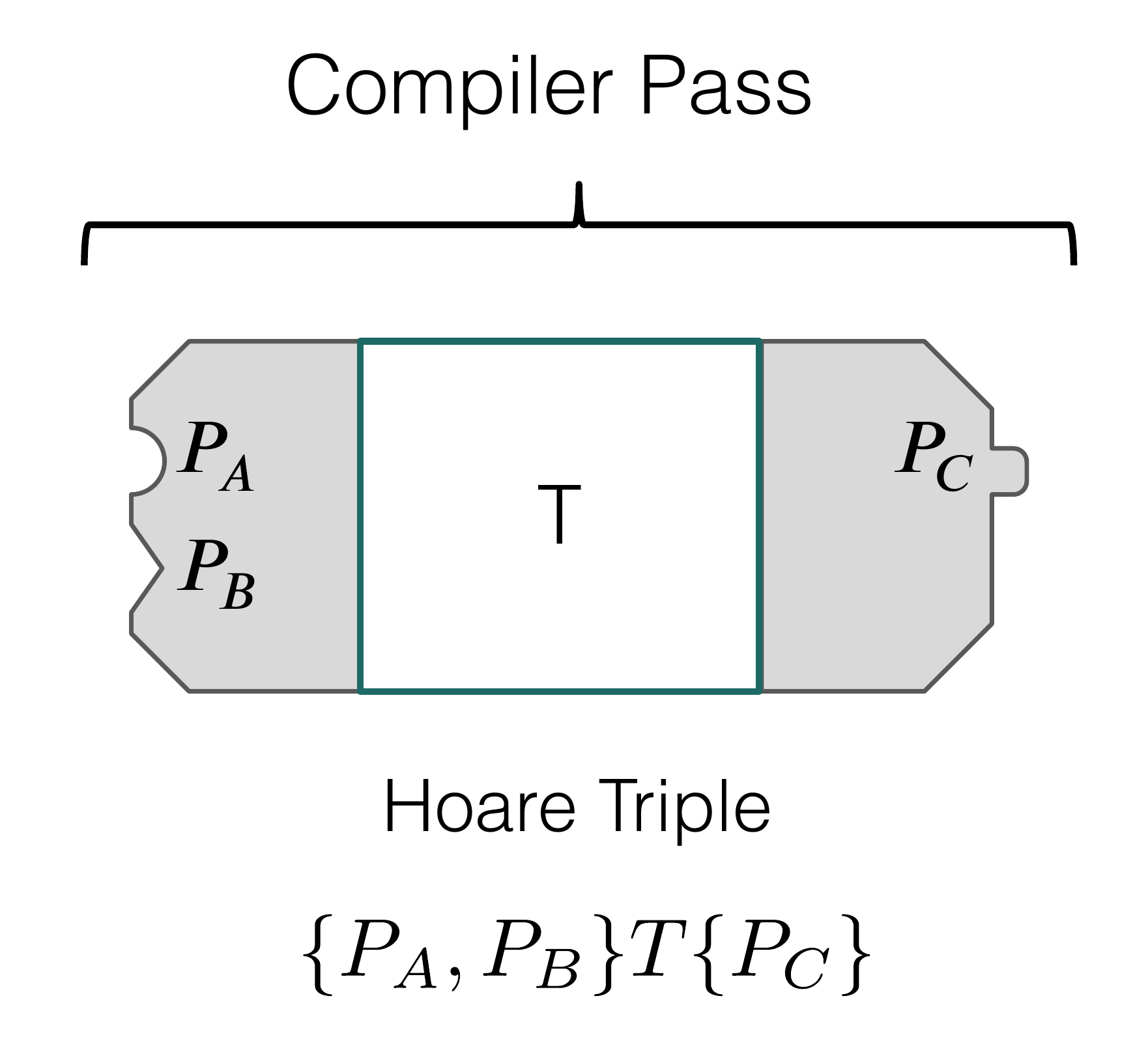}
  \end{center}
  \caption{A compiler pass is a transform pass with associated pre- and
  post-conditions.}
  \label{fig:pass_one}
\end{subfigure}

\begin{subfigure}{\textwidth}
  \begin{center}
  \includegraphics[scale=0.3]{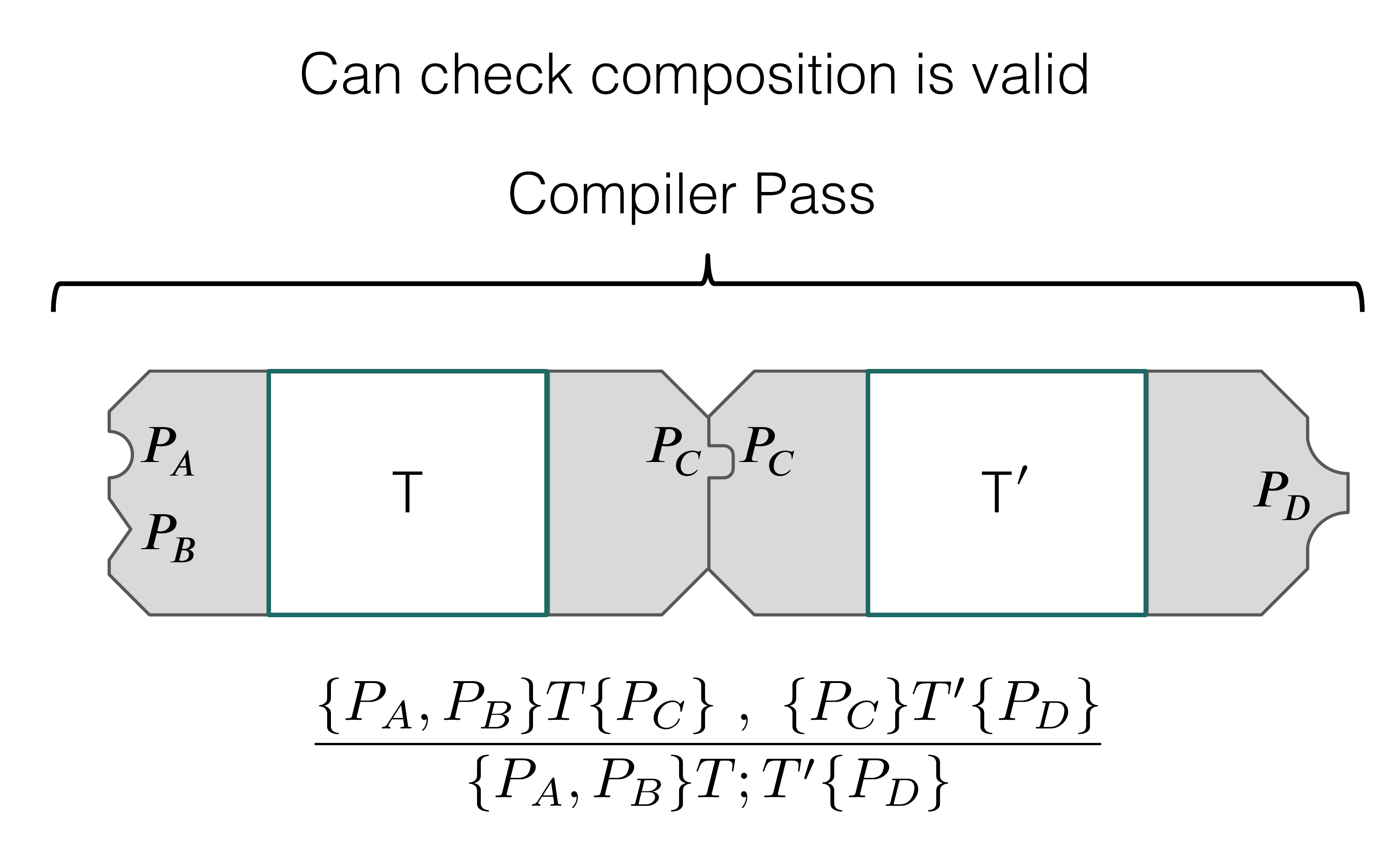}
  \end{center}
  \caption{Composition of compiler passes. The resulting Hoare triple
  is the standard sequential execution schema for two programs.}
  \label{fig:pass_two}
\end{subfigure}
\end{figure}

Each transform pass has a \emph{precondition} and a \emph{postcondition}, so
that the resulting compiler pass is a Hoare triple. This is illustrated in
Figure~\ref{fig:pass_one}. The compiler pass may be used on a circuit that
satisfies the precondition, and will guarantee that afterwards the circuit
satisfies the postcondition. Both the precondition and the postcondition are
sets of predicates. For example, a peephole optimisation may require that the
circuit be presented in a certain gate set before it can be applied. This
gate-set predicate forms the precondition of the pass. The optimisation can then
guarantee to the user that the rewrite rule will return the circuit in a
different gate set; this guarantee is the postcondition. 

These Hoare triples may then be composed, so that a custom rewriting
sequence can be generated, as shown in Figure~\ref{fig:pass_two}. If
the triples are correctly matched, so that no intermediate conditions
are conflicting, the custom sequence is valid.

More sophisticated combinators, such as loops, can be useful for
optimisation passes. For example, a user may wish to continue applying
a sequence of rewrite rules until no further rewrites are possible.
These combinators may be composed in the same way as sequences. When
looping combinators are used, termination of the resulting pass is not
guaranteed.

The full list of compiler passes and combinators can be found at
\url{https://cqcl.github.io/pytket/build/html/passes.html}.

\section{Circuit Optimisation Methods}
\label{sec:circ-optim-meth}

With the limited fidelity available on NISQ devices, effective circuit
optimisation is essential in order to extract all available performance out of
the machines. The goal is to identify equivalent circuits that will  accumulate
less noise when run on a real device.

Circuit optimisations in \tket are provided as compiler passes, which can be 
composed into larger routines. High performance is obtained by optimising at 
each stage in the compilation pipeline, so it is beneficial to have both 
powerful optimisations that can yield better results when not constrained by 
qubit connectivity or gate set and procedures targeted at specific 
architectures. \tket contains some methods that are architecture-agnostic and 
others that are architecture-aware (parametrised over the properties of the
device). Many of  the architecture-agnostic passes will additionally preserve
any connectivity  already satisfied by the inputs, allowing them to be applied
after routing.  Designing optimisations in this way provides retargetability
without  sacrificing performance.

\subsection{Circuit metrics}

Attempting to use the actual fidelity as a cost function would require accurate 
simulation of the quantum circuit with realistic noise models, which is both 
computationally expensive and highly dependent on a specific target 
architecture. Further, because real devices have noise sources that are complex
and hard to characterise, simple extrapolation from single-gate performance can
significantly overestimate the actual performance of the device, necessitating
more sophisticated, holistic
measures~\cite{Cross:2018aa,Blume-Kohout:2019ab,Erhard:2019aa}. However, simpler
metrics can give good, device-independent approximations to noise.

Naively optimising for gate count acknowledges the key fact that all gates will
introduce some degree of noise. However, NISQ devices tend to provide fast,
high-fidelity single-qubit rotations, with the error rates of multi-qubit 
operations being an order of magnitude worse~\cite{Arute:2019aa}. The primary 
focus for most optimisations in \tket is to minimise the \emph{two-qubit gate 
count}, which penalises the use of these slower and less accurate operations.

\begin{definition}
The \emph{two-qubit gate count} of a circuit is the number of
maximally-entangling two-qubit gates used in the circuit.
\end{definition}

This is often referred to as \CX-count, since any other maximally-entangling
two-qubit operation (such as \CZ or $e^{iXX\pi/4}$) is equivalent to a single
\CX up to local unitaries. This is analogous to the \emph{T-count metric} used 
at the error-correcting level.

Omitting single-qubit gates entirely from consideration improves 
device-independence, since the number of gates required varies significantly 
with the gate set (for example, a single U3 gate from the IBM specification can 
capture any rotation, while up to three are needed if decomposed into the 
underlying Rz and Ry gates).

The short coherence times of qubits strongly correlates the fidelity of the 
circuit with the time taken to execute. An ideal device will be able to 
parallelise gates acting on disjoint qubits to mitigate this. We can obtain a 
good approximation to the time taken on such a device by considering the
\emph{depth} of the circuit.

\begin{definition}
For a gate $g$ with predecessors $P(g)$, we define $\mathrm{depth}(g)$ by:

\begin{equation*}
\mathrm{depth}(g) \triangleq \left\{ \begin{array}{lr}
  0 & P(g) = \emptyset \\
  1 + \max_{p \in P(g)} \mathrm{depth}(p) & P(g) \neq \emptyset
  \end{array} \right.
\end{equation*}

The \emph{depth} of a circuit is the maximum value of $\mathrm{depth}(g)$ over 
all gates $g$. For any gate type $G$, the $G$-\emph{depth} of the circuit is
obtained by considering only the contribution from $G$ gates.
\end{definition}

Again, given the characteristics of multi-qubit operations on current hardware, 
\CX-depth (or depth with respect to any other maximal two-qubit gate) gives a
device-agnostic estimate of the time cost of a circuit.

\subsection{Peephole optimisations}

Circuit optimisations in \tket can broadly be categorised into \emph{peephole 
optimisations} and \emph{macroscopic analysis}. Peephole optimisations are
analogous to their namesake in classical compilers, where a sliding window
traverses the instruction graph, looking for specific small patterns or classes
of subcircuits and substituting equivalent subcircuits (with lower gate counts
or depth) in place. Basic examples include the elimination of redundant gates 
such as identities, gate-inverse pairs, and diagonal gates before measurements. 
Local gate commutation rules can be considered at the point of pattern 
identification, or as standalone passes to enable further optimisations.

These techniques are generic, in the sense that they are not tuned for
particular applications. The majority are written for best performance
in the intermediate gate set of \CX, Rz, and Rx, though when they can
be expressed more naturally in a different gate set (such as a set of
Clifford gates), rebase passes can be applied to convert between them.

\begin{figure}
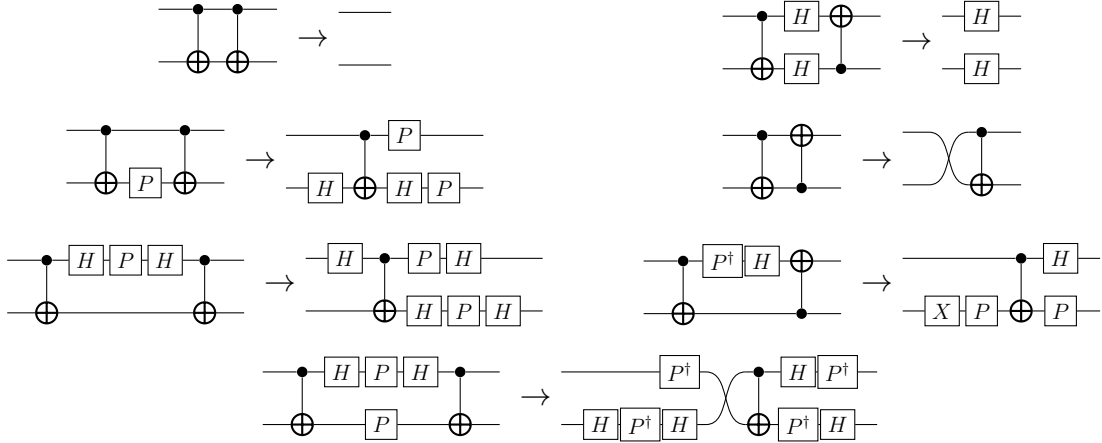

\begin{subfigure}[b]{0.5\textwidth}
\[
\inltf{Clifford0l} \to \inltf{Clifford0r}
\]
\vspace{0.4em}
\[
\inltf{Clifford1l} \to \inltf{Clifford1r}
\]
\vspace{0.4em}
\[
\inltf{Clifford2l} \to \inltf{Clifford2r}
\]
\end{subfigure}
\begin{subfigure}[b]{0.5\textwidth}
\[
\inltf{Clifford6l} \to \inltf{Clifford6r}
\]
\vspace{0.4em}
\[
\inltf{Clifford4l} \to \inltf{Clifford4r}
\]
\vspace{0.4em}
\[
\inltf{Clifford5l} \to \inltf{Clifford5r}
\]
\end{subfigure}
\vspace{0.6em}
\[
\inltf{Clifford3l} \to \inltf{Clifford3r}
\]
\caption{Clifford identities that can be recognised and applied in \tket to 
reduce the \CX count. $\mathrm{Rz}(\frac{\pi}{2})$ (phase) gates are represented 
by $P$ in the diagrams. The two identities that would introduce SWAPs can 
invalidate any connectivity, so can optionally be disabled.}
\label{fig:cliffordrules}
\end{figure}

Clifford circuits are defined as the class generated by \CX, Hadamard, and $\mathrm{Rz}
(\frac{\pi}{2})$ gates. These are known to be efficiently
simulable~\cite{Gottesman:1999to,Scott-Aaronson:2004yf}, and there is a wide
literature on simplification
techniques~\cite{amy2016finite,EPTCS287.5,selinger2013generators}.  In
particular, there are several useful small identities for reducing the 
\CX-count of a circuit, which \tket can recognise and apply: these are 
summarised in Figure~\ref{fig:cliffordrules}.

The Cartan decomposition~\cite{khaneja2001cartan} specifies a way to synthesise 
arbitrary $n$-qubit unitaries into sequences of local unitaries on fewer qubits 
and a small number of entangling operations between them. This decomposition
gives the following upper bounds for small instances:

\begin{theorem}
Any single-qubit unitary can be decomposed into a sequence of at most three 
rotations using any choice of Rx, Ry, and Rz gates. The angles of rotation are 
given by the Euler-angle decomposition of the combined rotation on the  Bloch
sphere.
\end{theorem}

\begin{theorem}
Any two-qubit unitary can be synthesised using at most three \CX gates and 15 
parametrised single-qubit gates (from any choice of Rx, Ry, and Rz), given by 
the $KAK$ decomposition~\cite{blaauboer2008analytical,vidal2004universal}.
\end{theorem}

\tket implements the Euler and $KAK$ decompositions by scanning the circuit
graph for long sequences of gates over one or two qubits and replacing them
whenever this helps to reduce \CX count or overall gate count. Known closed-form
expressions for manipulating Euler angles allow the single-qubit reduction to be
performed on symbolic circuits. \tket does not currently support performing the
$KAK$ decomposition with symbolic gate parameters, or a generic Cartan
decomposition for more than two qubits.

\subsection{Macroscopic optimisations}

Other optimisation procedures aim to identify high-level macroscopic
structures in the circuit or alternative mathematical representations
of different classes of circuits that are easier to manipulate than
individual gates. The general procedure here is to recognise these
structures or subcircuits of the appropriate class and treat them as
first-class gates. The algebra of the structure or representation can
identify non-local optimisations on the original circuit. Efficient
synthesis methods can then be applied to reduce these back down to
primitive gates in a way that uses fewer \CX gates, parallelises them
better, or restructures the circuit to enable more peephole
optimisations.

As simulation of molecular systems is a leading candidate application
for NISQ devices, \tket implements a novel technique for optimising a
new class of multi-qubit subcircuits, called \emph{Pauli gadgets},
which occur frequently in chemistry circuits designed for this
purpose. 

\begin{definition}
The \emph{phase gadget} $\Phi_n(\alpha)$ is a canonical representation of a 
multi-qubit operator of the form $e^{\frac12 i \alpha Z^{\otimes n}}$.
\[
\Phi_1(\alpha) :=  Z(\alpha)
\qquad\qquad
\Phi_{n+1}(\alpha) :=
(\CX \otimes 1_{n-1}) ;
(1_1 \otimes \Phi_n(\alpha)) ;
(\CX \otimes 1_{n-1})
\]
\[
\inltf{GadgetDef-lhs} 
\quad = \quad 
\inltf{GadgetDef-rhs} 
\]
\end{definition}

\begin{definition}
The \emph{Pauli gadget} $P(\alpha,s) := U(s) ; \Phi_{\sizeof{s}}(\alpha) ;
U(s)^\dag$ is a  canonical representation of a multi-qubit operator of the form
$e^{\frac12 i \alpha s}$, where $s$ is a string (tensor product) of Pauli 
operators and the unitary $U(s)$ is defined recursively:
\[
U(Z s') := I \otimes U(s')   \qquad
U(Y s') := X(\frac{\pi}{2})  \otimes U(s') \qquad
U(X s') := H  \otimes U(s')
\]
\end{definition}

\begin{example}
The simplest construction of a Pauli gadget is a single parameterised rotation
gate conjugated by a cascade of \CX gates and some single-qubit Clifford gates.
\[
P(\alpha, IXYZ) = \inltf{GadgetIXYZ}
\]
\end{example}

The authors have previously written a comprehensive account of Pauli
gadgets and their use in \tket~\cite{Cowtan:2019aa}. Such gadgets
enjoy a powerful equational theory, giving rules for commutation,
merging, and interaction with Clifford gates, which are easily proven
using the \zxcalculus~\cite{Coecke:2009aa}. By recognising these
structures in the input circuit, optimising the sequences of gadgets,
and efficiently transforming them back to a standard gate set, we can
achieve depth reductions greater than 50\%. See
Figure~\ref{fig:chempass} for a summary of results comparing this
technique in \tket to other compiler stacks for optimising  circuits
relating to electronic structure problems.

The first step when optimising with macroscopic structures is to identify good 
candidates in the circuit. It is obviously preferable to work with circuits 
that are already built from the structures to simplify this step. The integration of \tket with 
application software can make this possible by, for instance, allowing users to 
directly insert Pauli gadgets into the circuit using the corresponding box type.

Future versions of \tket will expand on this area of optimisations to cover 
other useful intermediate representations, including phase
polynomials~\cite{Nam:2018aa,Amy2014Polynomial-Time},
\zxdiagrams~\cite{Duncan:2019aa,EPTCS287.5,Kissinger:2019aa}, Clifford
tableaus~\cite{Scott-Aaronson:2004yf,Maslov2017Shorter-stabili}, and
linear-reversible  functions~\cite{Maslov2005Toffoli-network}.

\subsection{Example procedure}

Each of these methods gives rise to a compiler pass that can either be invoked
on its own or composed (as described in Section~\ref{sec:transform}) into more
effective routines.  \tket comes with some predefined passes combining several
of these optimisations. Each backend has a default compilation pass, which
guarantees (as far as possible) that the output will be  compatible with the
backend's hardware or simulator requirements; these passes include a small 
selection of the peephole optimisations for fast, basic gate reduction.

Figure~\ref{code:passcode} demonstrates how to compose the basic passes in 
\texttt{pytket}, recreating the effect of the pre-built \verb|SynthesiseIBM| pass. 
Starting with \verb|RebaseIBM| will decompose multi-qubit gates into a 
consistent gate set that is easier to manipulate. The \verb|RemoveRedundancies| 
pass covers a handful of optimisations based on removing different types of 
redundant gates. Applying commutation rules can potentially uncover more 
candidates for removal, so the \verb|simplify| routine is repeated until the 
gate count stops decreasing.

\begin{figure}[h]
  \pythonexternal[xleftmargin=.1\textwidth, xrightmargin=.1\textwidth]{code/passes.py}
  \caption{Code example showing how individual optimisation passes can be 
  composed into a more complex routine}
  \label{code:passcode}
\end{figure}

\begin{figure}
\fbox{
\begin{minipage}{\textwidth}
\begin{description}
\item[CliffordSimp] Recognises patterns from Figure~\ref{fig:cliffordrules} to 
reduce \CX count.
\item[CommuteThroughMultis] Commutes single-qubit gates through multi-qubit 
gates towards the front of the circuit.
\item[KAKDecomposition] Reduces two-qubit subcircuits to at most three \CX 
gates using the $KAK$ decomposition~\cite{blaauboer2008analytical,
vidal2004universal}.
\item[OptimiseCliffordsZX] Represents a Clifford circuit as a \zxdiagram, 
reduces it to a canonical form, and resynthesises it as a circuit.
\item[OptimisePhaseGadgets] Identifies phase gadgets~\cite{Cowtan:2019aa} in 
the circuit and resynthesises them in a shallow manner, attempting to align \CX 
gates between adjacent gadgets for further simplification.
\item[PauliSimp] Represents a circuit as a sequence of Pauli gadgets and a 
Clifford circuit, then resynthesises them pairwise~\cite{Cowtan:2019aa}.
\item[RemoveRedundancies] Removes redundant gates, including gate--inverse pairs,
identity rotations, diagonal gates before measurements, and adjacent rotations 
that can be merged.
\item[USquashIBM] Merges adjacent U1, U2, and U3 gates into a single U1 or U3 
gate by Euler-angle decomposition.
\end{description}
\end{minipage}}
\caption{Some of the elementary optimisation passes available in \tket}
\label{tab:passlist}
\vspace{5mm}
\end{figure}

\section{Mapping to Physical Qubits}
\label{sec:routing}

\begin{figure}[tb]
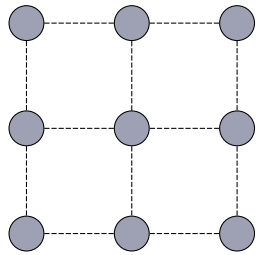

  \begin{center}
  \inltf{square_graph}
  \caption{An undirected graph showing connectivity constraints for a hypothetical 9 qubit device.}
  \label{fig:devicegraph}
  \end{center}
\end{figure}

Quantum computing devices have different constraints on possible
operations between their physical qubits. Some hardwares allow
two-qubit (or higher order) operations between any set of physical
qubits, while others do not. We define some physical qubits as being
connected if the hardware's primitive multi-qubit operations can be
executed between them. The connectivity constraints of a hardware can
be specified by an undirected graph \(G_D = (V_D, E_D)\), where the
vertices \(V_D\) are the physical qubits and edges \(E_D\) connect
physical qubits which can interact. Figure~\ref{fig:devicegraph} shows
an example connectivity graph.
As logical quantum circuits are usually written without considering
these connectivity constraints, they typically must be modified before execution on a hardware
to ensure that every logical multi-qubit operation is mapped to connected
physical qubits.
We define logical qubits and operations as those present in the logical quantum circuit
and state the routing problem as finding a mapping of logical operations
to allowed physical operations. 

\begin{figure}[htb]
  \begin{center}
  \includegraphics[scale=1.2]{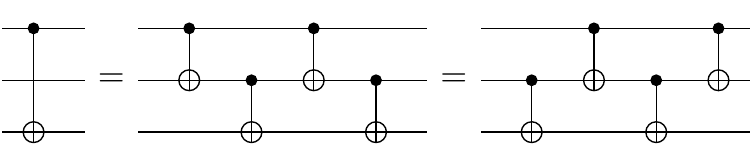}
  \caption{Distance 2 distributed \CX gate and decompositions to distance 1 \CX gates}
  \label{fig:bridge}
  \end{center}
\end{figure}

\begin{figure}[htb]
  \begin{center}
  \includegraphics[scale=1.2]{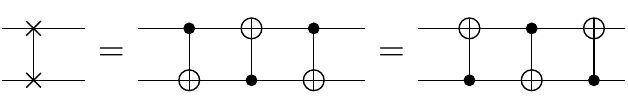}
  \caption{SWAP gate and decompositions to \CX gates}
  \label{fig:swap}
  \end{center}
\end{figure}

The routing problem is solved by permuting the mapping of logical
qubits to physical qubits throughout a circuit's execution, which is
achieved by adding SWAP operations. Sometimes gate decompositions can
be used to convert non-adjacent multi-qubit gates to distance 1
implementations. An example of this for the \CX gate is shown in
Figure~\ref{fig:bridge}. Finding an optimal solution to the routing
problem in this manner is NP-complete in general
\cite{Alexander-Cowtan:2019aa}. Note that SWAP operations have
different implementations on different hardwares; ion trap devices
have physical SWAPs while superconducting devices require the logical
states to be transferred between two physical qubits through three \CX
gates, shown in Figure~\ref{fig:swap}.
A solution is reached when the logical circuit is modified such that
there is an injective map of logical qubits to
physical qubits, or \textit{placement}, \(p\) where for every two-qubit gate
acting on logical qubits \((q, q')\), the mapped physical qubits are connected 
on the connectivity graph \(G_D\), or that \((p(q), p(q')) \in E_D\). 

In some cases a placement \(p\) can be found that solves the routing problem without adding SWAP operations.
Treating logical qubits as vertices and two-qubit interactions between them in
the circuit as edges, we can form an interaction graph for a logical circuit \(G_I = (V_I, E_I)\). 
If there is
a subgraph monomorphism \(p : V_I \to V_D\) which respects \((q, q') \in
E_I \Rightarrow (p(q), p(q')) \in E_D\), then only a relabelling of logical qubits
to physical qubits is required.

\begin{figure}[htb]
  \begin{center}
  \includegraphics[scale=1.2]{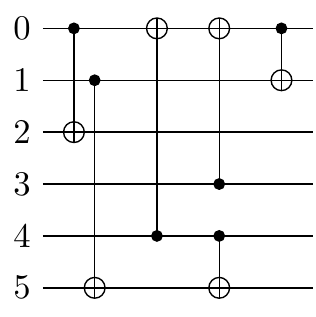}
  \caption{An example 6 qubit circuit with only \CX gates.}
  \label{fig:interactioncircuit}
  \end{center}
\end{figure}

In Figure~\ref{fig:interactioncircuit}, a short example \CX circuit is
shown. This circuit can be mapped to the connectivity graph from
Figure~\ref{fig:devicegraph} without the addition of additional gates,
as shown in Figure~\ref{fig:interactiongraph}.

\begin{figure}[htb]
  \begin{center}
  \inltf{interactiongraph}
  \caption{An example mapping of logical qubits in
  Figure~\ref{fig:interactioncircuit} to physical qubits in
  Figure~\ref{fig:devicegraph}. This example satisfies the routing
  problem without logical circuit modification. The solid lines
  between red nodes represent physical interactions performed by the
  circuit; grey nodes and dashed lines are unused by the circuit.}
  \label{fig:interactiongraph}
  \end{center}
\end{figure}

\tket~solves the problem in two
steps: finding an initial placement of logical qubits to physical qubits
and subsequent addition of SWAP operations
to the circuit. We consider this to be a \textit{dynamic} approach, in contrast 
to \textit{static} approaches \cite{Childs:2019aa,Zulehner:2018aa,Zulehner:2017aa}, that partition circuits
into parallelised slices of two-qubit interactions, and then use SWAP networks to
permute logical qubits between placements that satisfy these slices.\\

\subsection{Noise aware graph placement}
\label{sec:nagp}
The initial placement \(p\) is chosen to maximise the fidelity of the circuit
implementation on the device, using both a proxy heuristic which tries to minimise
additional gate overhead from routing, and an error heuristic which uses device
error characteristics. 

Indirectly, knowing only the connectivity graph \(G_D\), candidate placements
are chosen to minimise the number of gates the subsequent routing procedure will
need to add, as these additional operations are most likely error prone two
qubit gates. Routing will add gates dynamically as it proceeds through a
circuit, and so it is in general not possible to predict which placements will
correspond to the fewest gates added. As a heuristic, placements are found such
that a maximum number of two-qubit operations at the beginning of the circuit
can be completed with no SWAP gates added.

First this problem is cast as finding a
a subgraph monomorphism \(p : V_I \to V_D\) which respects \((q, q') \in
E_I \Rightarrow (p(q), p(q')) \in E_D\), for the interaction graph \(G_I = (V_I, E_I)\)
and device graph \(G_D = (V_D, E_D)\). 
If a monomorphism cannot be found, the
routine removes an edge from \(G_I\) belonging to the latest circuit slice
and attempts the graph matching again. This iterates until a monomorphism is
found. 

Logical qubits \(q \in V_I\) which no longer have any edges are 
removed from \(G_I\), thus the subgraph monomorphism routine in practice
produces a set of candidate \textit{partial} placements: maps which only act on
a subset of the logical qubits in the circuit. The subsequent routing procedure
can accept this as input, and will naively place unmapped qubits near those they
next interact with as it proceeds. As device architecture graphs \(G_D\)
currently have, and will likely continue to have, large regular subgraphs, the
set of matches can be large, especially when \(V_I\) is small compared to
\(V_D\).

If gate fidelity information for individual qubits is available for the target hardware,
these candidate placements are scored for maximum expected overall
fidelity and the highest scoring is chosen. In NISQ devices qubits and primitive
gates often have highly heterogeneous error characteristics, using this to
choose from the possible equivalent graph matches can result in a higher
fidelity implementation of a given circuit on a given device. Section 9.2
compares the performance of different placement methods available in \tket.

\subsection{Routing}
Given an intial partial placement \(p\), the routing algorithm 
adds SWAP operations until all logical operations satisfy connectivity constraints.
As SWAP operations are added, \(p\) is permuted, and so we define a temporary
placement \(p'\) which is the permutation of \(p\) from added SWAP operations
up to some slice of circuit gates \(S\).

Two-qubit gates in the circuit are iterated through in time order (via
a topological sort of the DAG), finding the first set of two-qubit
interactions \((q, q')\) such that \((p(q), p(q'))\) does not respect
\(G_D\) and no \(q\) is in multiple interactions. We call this set the
first slice \(S_0\) and log the permutation of \(p\), \(p'\), up to
\(S_0\). The routing algorithm then aims to pick the optimal edge \(e
\in E_D\) of the connectivty graph to implement a SWAP operation on,
given interacting logical qubits in \(S_0\) and \(p'\).

A set of candidate placements \(\{p''_{0},\dots,p''_{n}\}\) is
constructed by permuting instances of \(p'\) with SWAP operations on
edges in \(E_D\). If an edge has no qubits in \(S_0\) it is ignored.
Each candidate placement is scored and the winning placement is
chosen, with the scoring function based on the distance between
interactions in \(S_0\) given \(p''\). If there is no winning
placement for \(S_0\) then tied placements are scored for a new slice
\(S_1\), where \(S_1\) is the next set of set of two-qubit
interactions \((q, q')\) in the circuit such that \((p(q), p(q'))\)
does not respect \(G_D\) and no \(q\) is in multiple interactions. If
there is no winning placement for \(S_1\) then tied placements are
scored for a new slice \(S_2\). This is repeated until there is a
winning placement \(p''\) for some \(S_n\).

The winning placement \(p''\) is produced from \(p'\) via a permutation along its associated winning edge \(e\).
In most cases a SWAP operation is inserted along \(e\) directly before \(S_0\) and a new first slice \(S_0\) is found.

In some cases a distributed \CX is considered instead: at least one of the logical qubits associated with \(e\)
is in an interaction (a two-qubit gate \(g\)) in \(S_0\). 
If \(g\) is a \CX gate and its logical qubits are at distance two on the device graph \(G_D\) 
(for the temporary placement \(p'\)) then a distributed \CX may be added instead.
A new two element set of candidate placements is constructed comprised of \(p'\)
and \(p''\) and a similar scoring process is implemented, comparing \(p'\) and \(p''\) 
over multiple slices (\(S_0\), \(S_1\), \(S_2\) and so on). If \(p'\) wins \(g\) is replaced
with a distributed \CX and no SWAP operation is added. Else, if \(p''\) wins
the SWAP operation is added and \(p'\) replaced with \(p''\).

This whole process is then repeated, finding new first slices \(S_0\) and new winning placements \(p''\).
The algorithms terminates when \(S_0\) is returned empty.\\

The algorithm employs a high performance heuristic, which when coupled
with an efficient \cpluspluslogo\ implementation results in fast
runtime. \tket routing typically performs at least as well as other
software solutions when comparing circuit size and depth
\cite{Alexander-Cowtan:2019aa}.

Heterogeneity in NISQ device noise means a routed circuit with minimal
SWAP overhead may not always prove best though. Some solutions
consider device noise \cite{Murali:2019aa,Tannu:2018aa} when routing,
using gate fidelity information to produce a routed circuit with best
execution fidelity. This motivated a routing solution we implemented
that used a fidelity-based heuristic approach to score and pick SWAP
operations, for which the scoring method used an estimation of the
noise accumulated in realising all interactions in a slice \(S_0\).
The estimate was produced by finding SWAP paths required to permit
interactions in \(S_0\), and then calculating potential error accrued
by each logical qubit in realising these paths. In practice we found
that the fidelity heuristic could not accurately determine when
diverging from adding the minimal number of SWAP gates would improve
circuit fidelity, and so in general aiming to minimise SWAP operation
overhead provided better results. \\

\section{Applications}
\label{sec:applications}

Quantum chemistry simulations are performed using a supplementary software
package called Eumen. This provides an interface between traditional quantum
chemistry problems and various hybrid classical--quantum algorithms, enabling
effective chemistry simulations on NISQ hardware using \tket. For such
simulations, Eumen accepts a range of input parameters, such as the molecular or
lattice geometry, system charge, multiplicity restrictions, type of simulation,
ansatz, optimisations, hardware backend, and qubit mapping. \tket mediates
between Eumen and the hardware on which the quantum-algorithmic part of the
chemistry simulation runs.

Eumen can compute optimal geometries and properties of the ground state or
excited states. For example, ground-state energies may be calculated using VQE
or imaginary-time evolution methods. For excited-state calculations, one can use
methods such as quantum subspace expansion, reduced density matrix
approximation, penalty functions, and symmetry
constraints~\cite{PhysRevA.95.042308}. These methods require the measurement of
either the expectation value of many-body operators or the overlap of two
different states; these measurements are performed by \tket using results from
the quantum hardware. The states may be prepared with hardware-efficient
ans{\"a}tze or the approximated circuit representation of various physically
motivated ans{\"a}tze, such as UCCSD, k-UpCCGSD, or the time evolution operator.

The depth of the state preparation circuit is significantly reduced by \tket's
ansatz-specific optimisation methods, which can identify specific structures in
circuits and reduce the gate count required for their execution, as described in
Section~\ref{sec:circ-optim-meth}. Future optimisation methods specific to QAOA
instances are also planned. Construction of such structures and subsequent
optimisation is aided by boxes: for example, a Pauli gadget can be added to a
circuit in abstract form via a \texttt{PauliExpBox} operation. These operations,
and variational circuits in general, can make use of symbolic parameters and
compilation to simplify their use and speed up compilation. Finally, variational
algorithms, and other applications that use Hamiltonian estimation via
calculation of multiple terms of the Hamiltonian, can benefit from \tket's
back-end methods, which allow compilation and submission of multiple circuits.
Circuit execution on the back-end can occur asynchronously, with results being
retrieved for processing when execution is complete.

\section{Benchmarks}
\label{sec:benchmarks}

In this section we provide some benchmarks of \tket compiler performance. First
we conduct benchmarks of end-to-end compiler performance, including comparisons
with other available quantum compilation tools. Secondly, we perform experiments
on a publicly available quantum device to determine whether noise-aware
placement offers a benefit. The full datasets and scripts used for generating these results can
be found at \url{https://github.com/CQCL/tket_benchmarking}.

\subsection{End-to-end compilation}

We present a series of benchmarks of end-to-end compiler performance on a set of
circuits, and compare the performance of \tket with that  of two widely-used
alternatives, Qiskit and the Quilc compiler, which are able to do both general
circuit optimisation and routing. 

We define end-to-end compilation as the process of translating a circuit,
presented in OpenQASM, and outputting an equivalent circuit that has been
optimised and has the relevant device constraints satisfied, i.e.~has been
routed and converted to the correct gate set. We do not include high-level
algorithm design or low-level pulse optimisation which are respectively 
beyond the scope of a compiler and only in its infancy as a research topic.

\subsubsection{Benchmark circuits}

The benchmark set is made up of three test sets, with circuits of at most $10^4$
initial gates. (This threshold was chosen as circuits larger than this
are several orders of magnitude too large for near-term devices. In addition,
the runtimes of Qiskit and Quilc already reach several minutes per circuit at 
this range, making it impractical to benchmark against the larger circuits.)
\begin{enumerate}
  \item The IBM test set is a series of circuits published as part of the Qiskit
Developer Challenge, a public competition to design a better routing algorithm.
These circuits are not amenable to significant peephole optimisation to 
restrict the impact this can have and focus on the efficiency of the routing 
algorithm. However, they were designed to be easily verified for correctness by 
mapping the $\ket{0}^{\otimes n}$ state into some other computational basis 
state; as such, these tests could be circumvented by applying state preparation 
optimisations.
The IBM circuit set in OpenQASM can
be found at \url{https://github.com/iic-jku/ibm_qx_mapping}.
  \item The UCCSD test set is a series of circuits for electronic-structure
calculations. They correspond to VQE circuits for estimating the ground state
energy of small molecules by the Unitary Coupled Cluster
approach~\cite{Romero_2018}, using some choice of qubit encoding
(Jordan--Wigner, parity mapping, or Bravyi--Kitaev~\cite{Steudtner_2018}). These
circuits are very amenable to optimisation, as well as requiring routing. They
are representative of algorithms that have been proposed as suitable for
application on NISQ devices~\cite{Romero_2018}, and were generated using Qiskit
Aqua. The set used here updates and extends that used by Cowtan 
\etal~\cite{Cowtan:2019aa}, whose OpenQASM files can be found at
\url{https://github.com/CQCL/pytket}.
  \item The Product Formula test set is a series of circuits for Hamiltonian
simulation. These circuits are thought to be candidates for quantum
advantage~\cite{Childs:2018aa}, and were used as a test case for the circuit
optimizer by Nam \etal~\cite{Nam:2018aa}. They are given in the ASCII format of
the Quipper language, and each is formed of a repeated subroutine. We convert
this subroutine to a quantum circuit in OpenQASM. We included
circuits both before and after optimisation by Nam \etal~\cite{Nam:2018aa}, 
since they still require mapping to the architecture and have potential for 
further optimisation. We had to further edit the circuits to ensure the 
rotation angles of gates exceeded Qiskit's very high cutoff ($10^{-5}$), below 
which the rotations would be treated as identities, making these circuits 
almost trivial.
These circuits contain some Pauli gadgets, but also have large regions which are
not amenable to this kind of optimisation.  These circuits can be found at
\url{https://github.com/njross/optimizer}.
\end{enumerate}

The full collated benchmark set can be found at
\url{https://github.com/CQCL/tket_benchmarking}.

\subsubsection{Experiments}

\begin{figure}[tb]
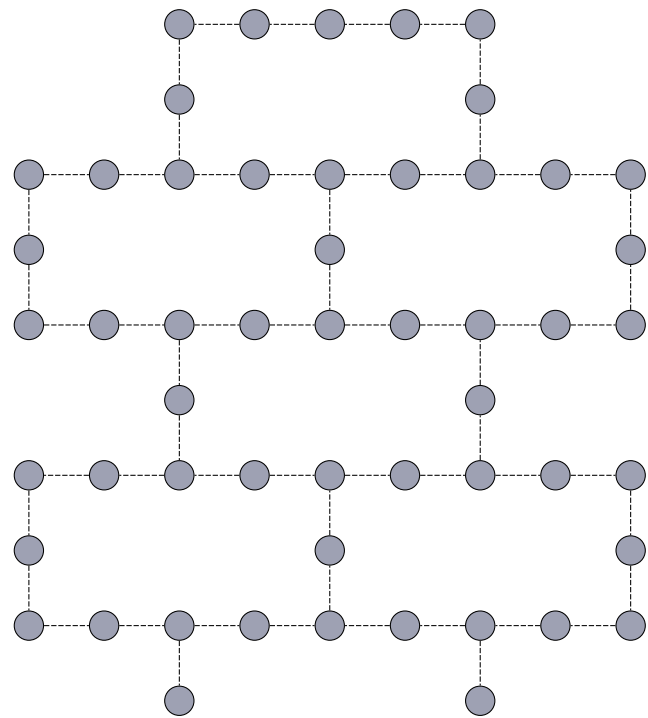
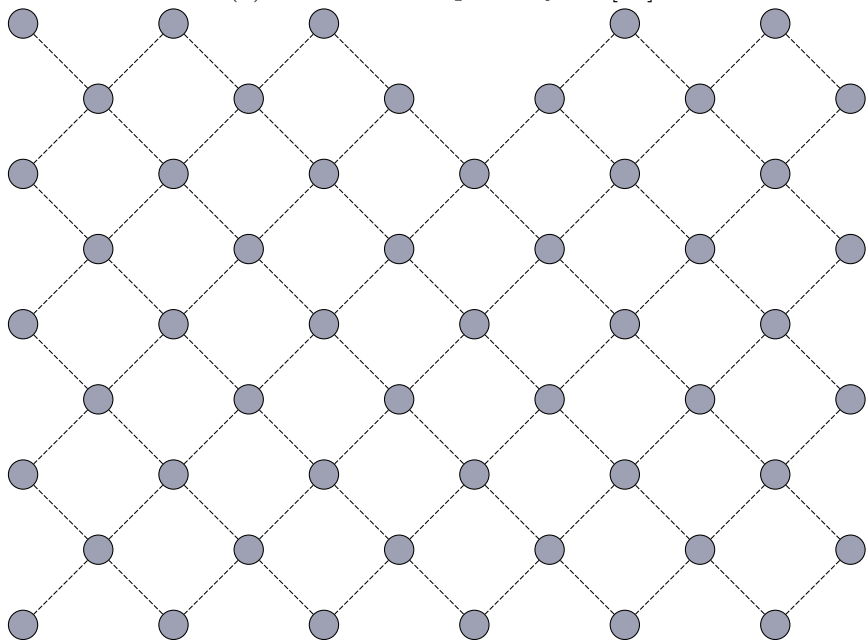
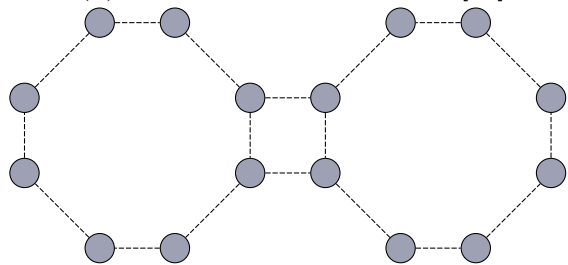

  \centering
  \begin{subfigure}[b]{\textwidth}
      \centering
      \inltf{rochester}
      \caption{Rochester 53-qubit layout \cite{QuantumComputingReport}.}
      \label{fig:rochester}
  \end{subfigure}

  \begin{subfigure}[b]{\textwidth}
      \centering
      \inltf{sycamore}
      \caption{Sycamore 53-qubit layout \cite{Arute:2019aa}.}
      \label{fig:sycamore}
  \end{subfigure}

  \begin{subfigure}[b]{\textwidth}
      \centering
      \inltf{aspen}
      \caption{Rigetti Aspen 16-qubit layout \cite{rigetti_blog}.}
      \label{fig:rigetti16q}
  \end{subfigure}

     \caption{Device connectivity layouts used in end-to-end compilation
     benchmarks.}
     \label{fig:arcs}
\end{figure}

We compare compilation for four different architectures:
\begin{enumerate}[label=\arabic*)]
\item The fully-connected graph, for which no routing is required.
\item The connectivity graph of IBM Rochester, a 53-qubit device.
\item The connectivity graph of Google's 53-qubit Sycamore device.
\item The Rigetti Aspen 16-qubit architecture. For this case, all
circuits with more than 16 qubits were discarded.
\end{enumerate}
These connectivity graphs are shown in Figure~\ref{fig:arcs}.
Two end-to-end comparisons were made:
\begin{enumerate}
  \item To compare \tket to other available compilation software, the default compiler passes of Qiskit (optimisation level 3) and Quilc, and the recommended generic pass for \tket (\texttt{FullPeepholeOptimise}, followed by the default qubit mapping pass, \texttt{SynthesiseIBM}, and the rebase pass into the desired gateset, henceforth referred to as `FullPass') were applied to all
available circuits.
  \item To demonstrate the necessity of appropriate usage of situational compiler optimisations, we compare \tket's UCCSD-specific pass (the \texttt{PauliSimp} pass followed by the `FullPass' routine, henceforth dubbed `ChemPass') against the default Qiskit
and Quilc passes, on the UCCSD circuits (test set~(ii) above). These circuits contain adjacent Pauli gadgets, and we
demonstrate that the reduction in two-qubit gate count and depth can be
substantial compared to optimising naively.
\end{enumerate}

The benchmarks were performed using \tket~v0.4.1, Quilc~v1.16.3 and
Qiskit Terra~v0.12.0. All results were obtained using a machine with a
2.3~GHz Intel Core i5 processor and 8~GB of 2133~MHz LPDDR3 memory,
running MacOS Mojave~v10.14.

\subsubsection{Metric}

The figures of merit for these benchmarks are two-qubit gate count and
depth. As described in Section~\ref{sec:circ-optim-meth}, two-qubit
gates have error rates an order of magnitude higher than single-qubit
gates for existing architectures~\cite{Arute:2019aa}, and so the
counts and depths are reasonable proxies for the overall expected
error rates of a circuit run on a NISQ device.

We defined end-to-end compilation earlier, including the conversion to
the device's native gate set. Google's
Sycamore device can accept \CZ gates natively as a two-qubit
operation, whereas IBM Rochester only supports \CX gates. 
As only single-qubit Hadamard
gates are required for conversion between \CX and \CZ gates, we
discount the gate-conversion step, and accept either gate set for the
two-qubit gate-count and depth metrics. Thus, unlike total gate count,
for these backends the two-qubit gate count should be independent of
final basis set chosen, meaning the comparison between architectures is purely 
based on connectivity. The exception here is the Rigetti Aspen device which can 
use both \CZ gates and the XY family (including the iSWAP gate) natively. Since 
these can be obtained with similar fidelities, the simple two-qubit gate count 
is justified in weighting their costs equally.

\subsubsection{Results}
\label{sec:end_to_end_results}
\begin{figure}[h]
  \includegraphics[width=\linewidth]{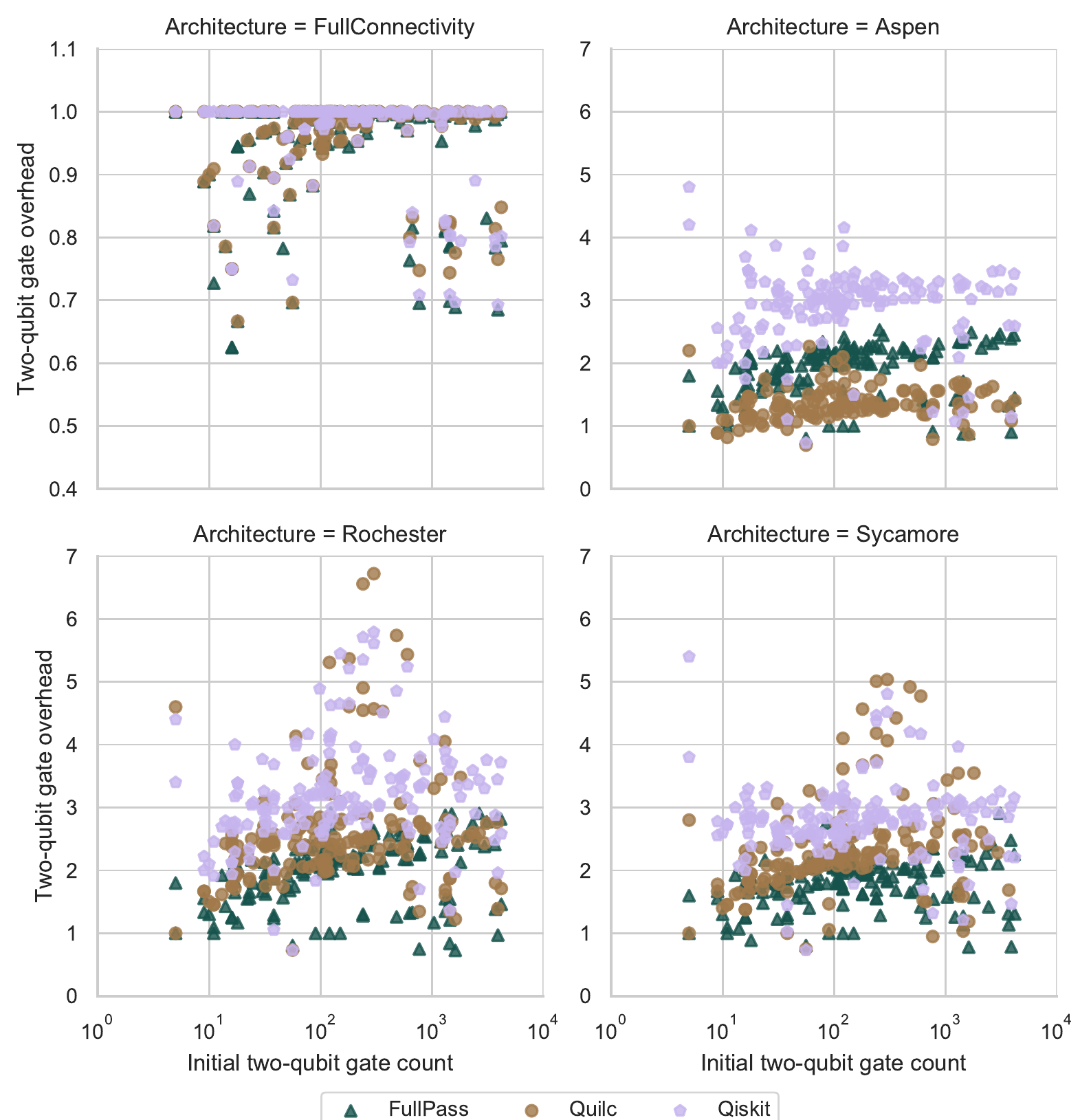}
  \begin{center}
    
    \begin{tabular}{@{}lllll@{}}
      \toprule
      & FullConnectivity & Aspen         & Rochester       & Sycamore        \\ \midrule
      FullPass  & $0.939\pm0.011$  & $1.864\pm0.039$ & $1.975\pm0.042$ & $1.773\pm0.034$ \\
      Quilc     & $0.949\pm0.010$  & $1.362\pm0.022$ & $2.595\pm0.074$ & $2.343\pm0.061$ \\
      Qiskit & $0.958\pm0.010$  & $2.896\pm0.052$ & $3.201\pm0.065$ & $2.789\pm0.047$ \\ \bottomrule
    \end{tabular}
  \end{center}
  \caption{
    Default compilation benchmarks over all benchmark circuits. The
    multiplicative overhead in two-qubit gate count from input circuit to output
    circuit is plotted against input two-qubit gate count. The table shows means
   across the circuit sets and associated standard error. In general, routing
    induces a larger-than-$1$ overhead, but for \textit{FullConnectivity} when no routing
    is required it is $1$ or below. `FullPass' refers to the recommended \tket routine, consisting of the
    \texttt{FullPeepholeOptimise} pass, followed by the corresponding qubit mapping pass, \texttt{SynthesiseIBM} and rebasing to the final gate set.
    }
  \label{fig:fullpass}
  
\end{figure}

\begin{figure}[h]
  \includegraphics[width=\linewidth]{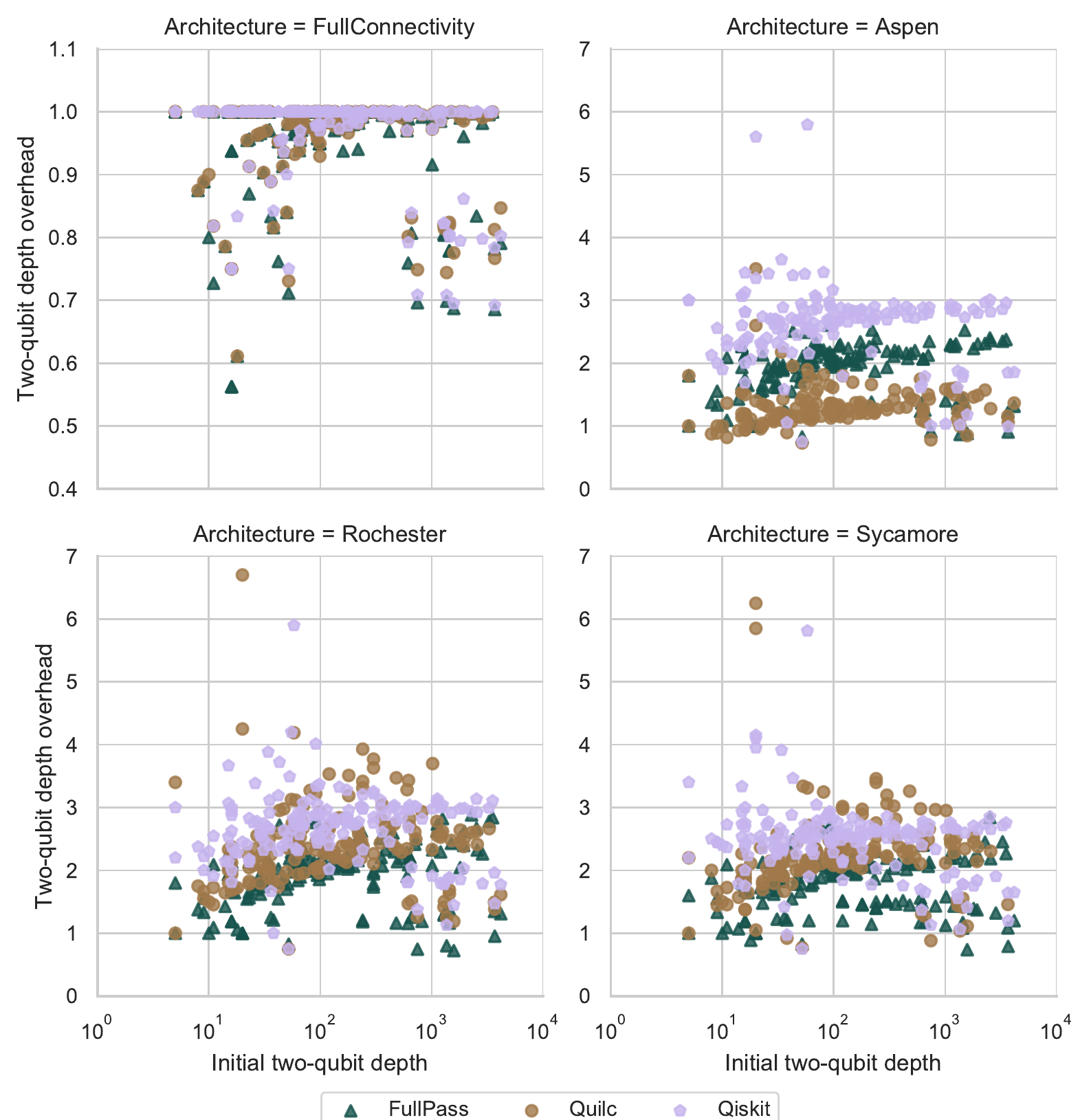}
  \begin{center}
    \begin{tabular}{@{}lllll@{}}
      \toprule
                & FullConnectivity & Aspen         & Rochester       & Sycamore        \\ \midrule
      FullPass  & $0.937\pm0.011$  & $1.844\pm0.040$ & $1.924\pm0.041$ & $1.780\pm0.038$ \\
      Quilc     & $0.950\pm0.010$  & $1.312\pm0.027$ & $2.418\pm0.063$ & $2.223\pm0.053$ \\
      Qiskit & $0.955\pm0.010$  & $2.631\pm0.065$ & $2.748\pm0.081$ & $2.473\pm0.045$ \\ \bottomrule
      \end{tabular}
    
  \end{center}
  \caption{
    Default compilation benchmarks over all benchmark circuits. The
    multiplicative overhead in two-qubit gate depth from input circuit to output
    circuit is plotted against input two-qubit depth. The table shows means
    across the circuit sets and associated standard error. In general, routing
    induces a larger-than-$1$ overhead, but for \textit{FullConnectivity} when no routing
    is required it is $1$ or below. `FullPass' refers to the recommended \tket routine, consisting of the
    \texttt{FullPeepholeOptimise} pass, followed by the corresponding qubit mapping pass, \texttt{SynthesiseIBM} and rebasing to the final gate set.
    }
  \label{fig:fullpass_d}
\end{figure}

\begin{figure}[h]
  \includegraphics[width=\linewidth]{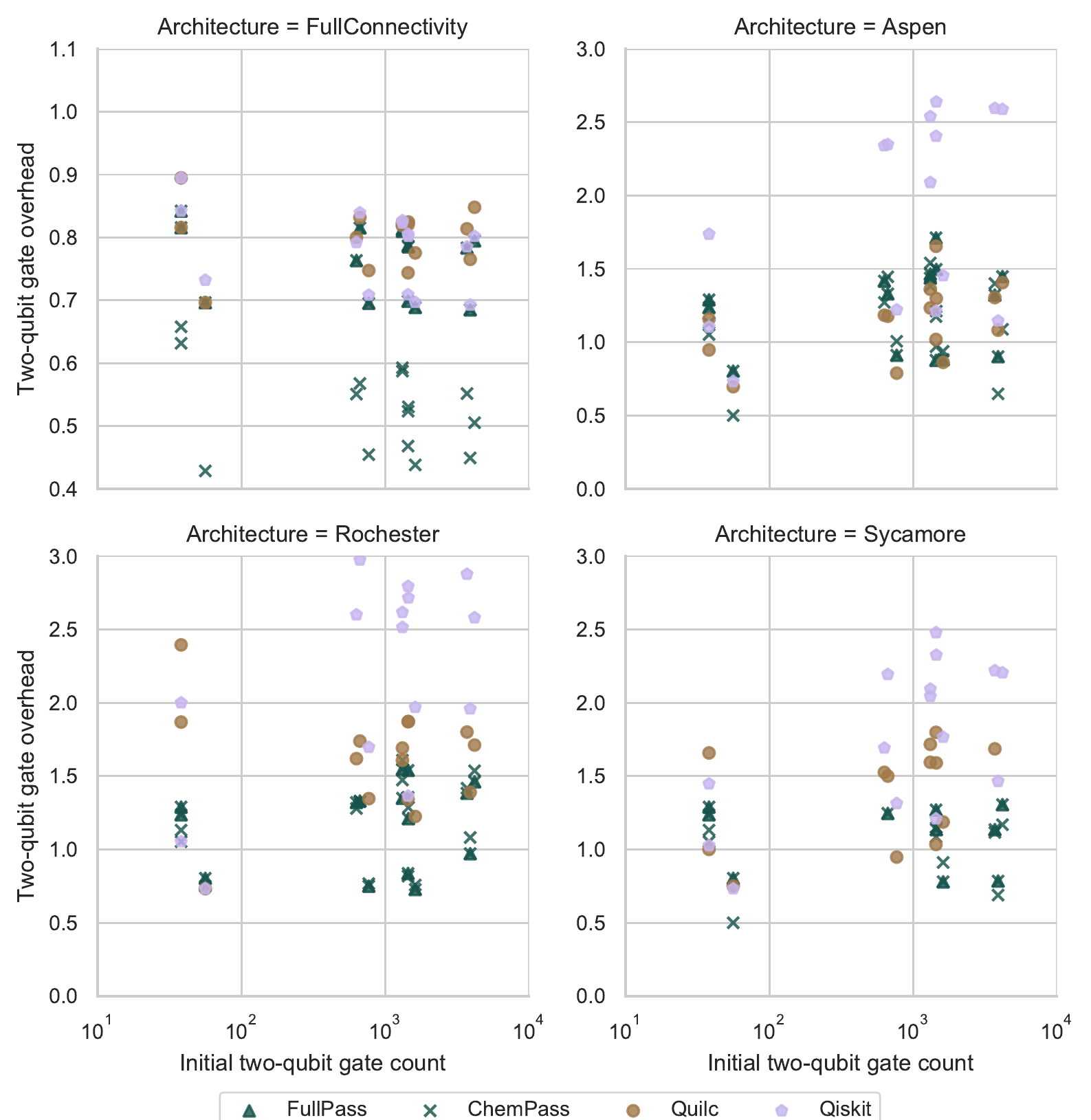}
  \begin{center}
    \begin{tabular}{@{}lllll@{}}
      \toprule
                & FullConnectivity & Rigetti         & Rochester       & Sycamore        \\ \midrule
      FullPass & $0.765\pm0.014$ & $1.236\pm0.074$ & $1.184\pm0.074$ & $1.099\pm0.070$ \\ 
      ChemPass  & $0.647\pm0.025$  & $1.176\pm0.052$ & $1.180\pm0.052$ & $1.042\pm0.053$ \\
      Quilc     & $0.801\pm0.013$  & $1.145\pm0.065$ & $1.614\pm0.098$ & $1.384\pm0.097$ \\
      Qiskit & $0.783\pm0.016$  & $1.877\pm0.173$ & $2.163\pm0.180$ & $1.747\pm0.136$ \\ \bottomrule
    \end{tabular}
  
  \end{center}
  \caption{
    Chemistry-specific compilation benchmarks over the UCCSD test set. The
    multiplicative overhead in two-qubit gate count from input circuit to output
    circuit is plotted against input two-qubit gate count. The table shows means
    across the circuit sets and associated standard error. `ChemPass' refers to
    application of the \tket \texttt{PauliSimp} pass, followed by the `FullPass' routine.
    }
  \label{fig:chempass}
\end{figure}

\begin{figure}[h]
  \includegraphics[width=\linewidth]{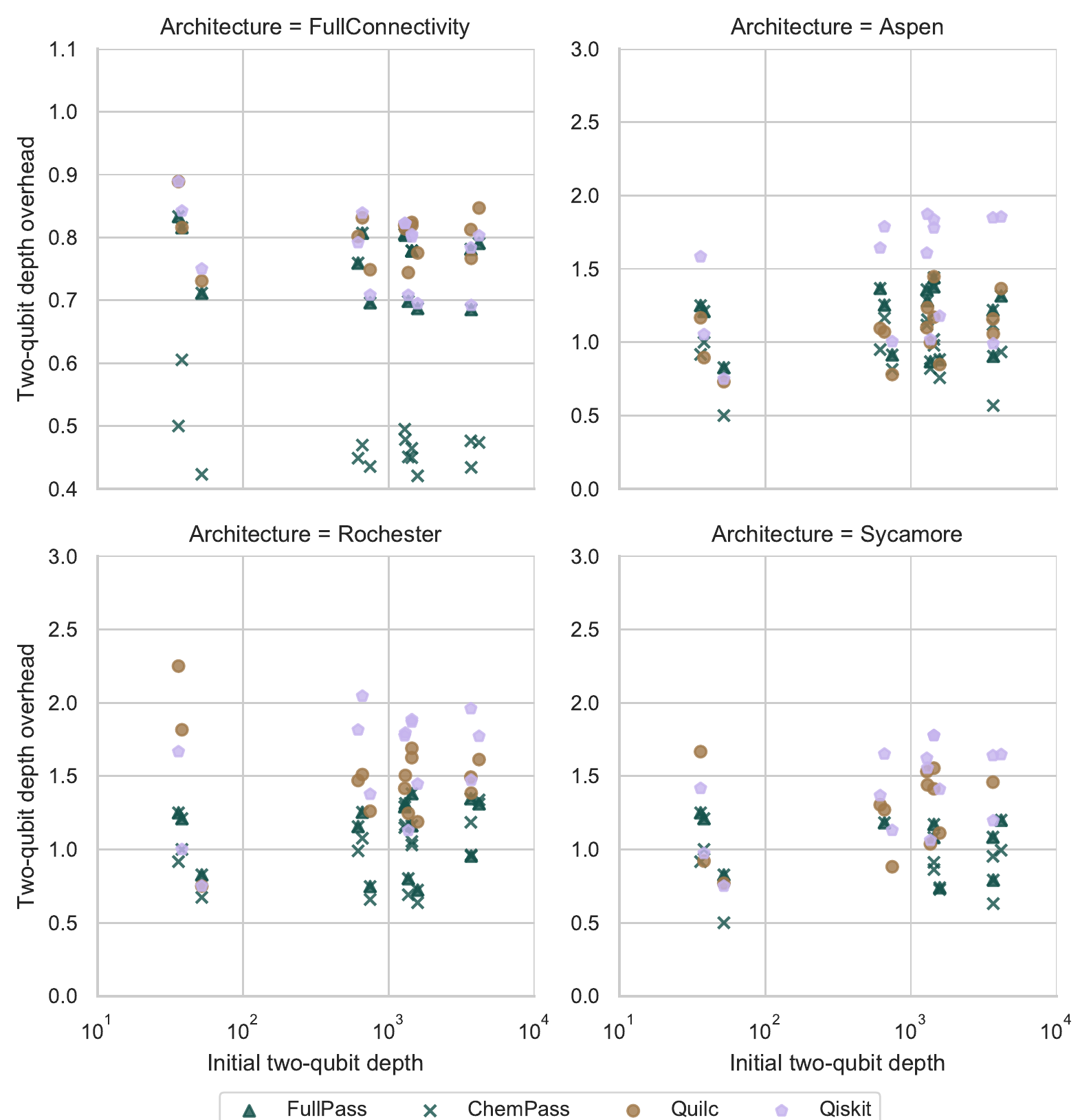}
  \begin{center}
    \begin{tabular}{@{}lllll@{}}
      \toprule
                & FullConnectivity & Rigetti         & Rochester       & Sycamore        \\ \midrule
      FullPass & $0.762\pm0.013$ & $1.165\pm0.056$ & $1.115\pm0.061$ & $1.054\pm0.061$ \\
      ChemPass  & $0.615\pm0.029$  & $1.043\pm0.044$ & $1.042\pm0.043$ & $0.949\pm0.049$ \\
      Quilc     & $0.803\pm0.011$  & $1.074\pm0.052$ & $1.481\pm0.085$ & $1.258\pm0.080$ \\
      Qiskit & $0.783\pm0.016$  & $1.454\pm0.104$ & $1.584\pm0.098$ & $1.398\pm0.081$ \\ \bottomrule
      \end{tabular}
  
  \end{center}
  \caption{Chemistry-specific compilation benchmarks over the UCCSD
    test set. The multiplicative overhead in two-qubit depth from
    input circuit to output circuit is plotted against input two-qubit
    depth. The table shows means across the circuit sets and
    associated standard error. `ChemPass' refers to application of the
    \tket \texttt{PauliSimp} pass, followed by the `FullPass'
    routine.}
  \label{fig:chempass_d}
\end{figure}

Two-qubit gate count
and depth benchmark results for default compilation are shown in
Figures~\ref{fig:fullpass} and~\ref{fig:fullpass_d}. The corresponding results
for chemistry-specific compilation are shown in Figures~\ref{fig:chempass}
and~\ref{fig:chempass_d} respectively.
Each figure includes results for all three compilers and all four target backends.
The figure of merit in each is multiplicative overhead of two-qubit gate count
or depth, i.e. the ratio of the values before and after compilation.

The \textit{FullConnectivity} case shows the results without the effects of
routing. For default compilation passes the majority of benchmark circuits show
little to no change in the number of two-qubit gates, demonstrating the
difficulty of entangling-gate reduction with generic optimisation passes. In
cases where gains can be made, the optimisations in \tket are able to make
larger reductions than the other compilers, and are able to make reductions in
more instances.

By contrast, as shown in Figures~\ref{fig:chempass} and
\ref{fig:chempass_d}, adding the \texttt{PauliSimp} pass leads to more
significant reductions on the UCCSD circuits, even after mapping onto
a device with restricted connectivity. However, on other classes of
circuits that don't resemble the UCCSD set, adding this optimisation
can cause a drastic drop in performance, by trying to fit them to this
model, and potentially making them less amenable to routing.

Across the end-to-end compilation results targeting
restricted-connectivity architectures, the general ranking of
performance is: \tket, followed by Quilc (with special note of their 
performance for the Aspen device), with Qiskit consistently introducing 
a very high gate overhead. The results are sufficiently spread that for
circuits viable for execution on near-term devices, the choice of
compiler and pass sequence makes a significant difference to the size of the final
circuit. By comparing to the \textit{FullConnectivity} case, we can
see that these differences are dominated by differences in routing
performance.

\subsection{Noise-aware placement}
\label{sec:noise-aware-plac}

In Section~\ref{sec:nagp} we outlined \emph{graph placement} (GP), a
subgraph-monomorphism-based method for finding initial qubit mappings, and
\emph{noise-aware graph placement} (NAGP), a fidelity-aware heuristic for
scoring those placements. As the effectiveness of these methods depends strongly
on the error characteristics of physical devices and the fidelity with which
they execute the circuit, we assess and compare their performance by running
benchmark circuits on a device with and without the mapping methods applied. 

\subsubsection{Benchmark circuits}

At the time of writing, it is difficult to implement many common algorithms on
publicly-available quantum devices and extract a signal from the noise; this is
why implemented algorithms are usually variational. In order to test a large
enough data set to have confidence in measured differences, we instead choose to
implement random circuits of constrained sizes.

Sets of random circuits are parameterised by number of qubits and
total number of gates. Each circuit is generated by uniformly sampling
gates from \(\{\mathrm{X}, \mathrm{Y}, \mathrm{Z}, \mathrm{H},
\mathrm{T}, \mathrm{S}, \mathrm{CX} \}\), and uniformly sampling from
all qubits for single-qubit gates and from the set of pairs of all
qubits in the case of a \CX. As the two-qubit operations are the most
error-prone~\cite{Arute:2019aa}, samples that included no \CX gates
were excluded. Circuit sets of qubit numbers of 4 and 8, and gate
counts of 20, 40, 60 and 80, were generated. Circuits over 4~qubits
with 80~gates were too deep and therefore noisy for effective
comparison of methods, so were omitted, leaving a total of 7~sets each
with 90~samples.

\subsubsection{Experiments}

As described in Section~\ref{sec:routing}, \emph{placement} is the task of
finding  initial maps from logical qubits of the circuit to physical qubits of
the device, and \emph{routing} is the addition of two-qubit gates to satisfy the
connectivity constraints of the device. For these experiments, each benchmark
circuit was compiled in three different ways, corresponding to three methods of
calculating an initial partial placement: ``None'' (corresponding to no qubits
placed, therefore relying on default on-the-fly placement performed by routing),
``Graph Placement'', and ``Noise-Aware Graph Placement''. Each placed circuit
was then compiled with identical routing and post-routing optimisation passes
from \tket.

All circuits were run on the publicly-available \texttt{ibmq\_16\_melbourne}
device via the IBM~Q Experience~\cite{ibmq}. Correspondingly, compilation also
included translating the circuits to the IBM~Q gate set of \(\{\mathrm{U_1},
\mathrm{U_2}, \mathrm{U_3}, \mathrm{CX}\}\). Programs for execution on an IBM~Q
device are sent via the API as ``jobs'', with a maximum of 75~circuits in each
job. All compiled circuits that corresponded to the same initial circuit were
evaluated consecutively and within the same job, to mitigate the effects of
device-characteristic deviations between jobs on the method comparisons. Each
compiled circuit was evaluated with the maximum 8192~shots.

\subsubsection{Metric}

The ideal metric for comparing the placement methods is overall fidelity of the
implemented circuits. However, measurement of this involves a number of circuit
measurements that scales exponentially with qubit number and is infeasible for
large experiment sets~\cite{Flammia:1346875}. Instead, we choose a metric that
can quantify the distance between the distribution of measurements in the
computational basis and the same distribution generated from an ideal
simulation. This methodology requires classical resources that scale
exponentially with qubit number, as it involves simulation of all circuits.
However, as the techniques proposed are only relevant for NISQ-era error rates
and device heterogeneity, we expect the number of qubits to remain low enough
for applicability of the proposed methods while the same levels of heterogeneity
also hold.

The Kullback--Leibler (KL) divergence has been used for comparing measured and
simulated distributions~\cite{Nishio:2019aa}. However, it has some shortcomings.
The KL divergence between two distributions \(P, Q\) over values \(x_i\) is
defined as: \[D_\mathrm{KL}(P, Q) = \sum_i P(x_i)\log\frac{P(x_i)}{Q(x_i)}\]
This is asymmetric between \(P\) and \(Q\). More importantly, it is defined to
be infinite if \(\text{support}(P) \nsubseteq \text{support}(Q)\).  A standard
technique to account for this is padding zeroes in the distribution with small
values and renormalising. Although this would still show qualitative difference,
the absolute value would depend on the free parameter of tuning, so would not be
a useful quantitative measure.
 
We instead use the Jenses--Shannon divergence
\(D_\mathrm{JS}\)~\cite{PhysRevA.72.052310}, which is closely related to
\(D_\mathrm{KL}\). For distributions \(P, Q\) and \(M = \frac{1}{2}(P + Q)\) it
is defined by:
\[
  D_\mathrm{JS}(P, Q) = \frac{1}{2}D_\mathrm{KL}(P, M) + \frac{1}{2}D_\mathrm{KL}(Q, M) 
\]

This is a symmetric, finite function with bounds \(0 \leq D_\mathrm{JS}(P, Q) \leq 1\)
when the base-$2$ logarithm is used.

\subsubsection{Results}

\begin{figure}[h]
  \includegraphics[width=\linewidth]{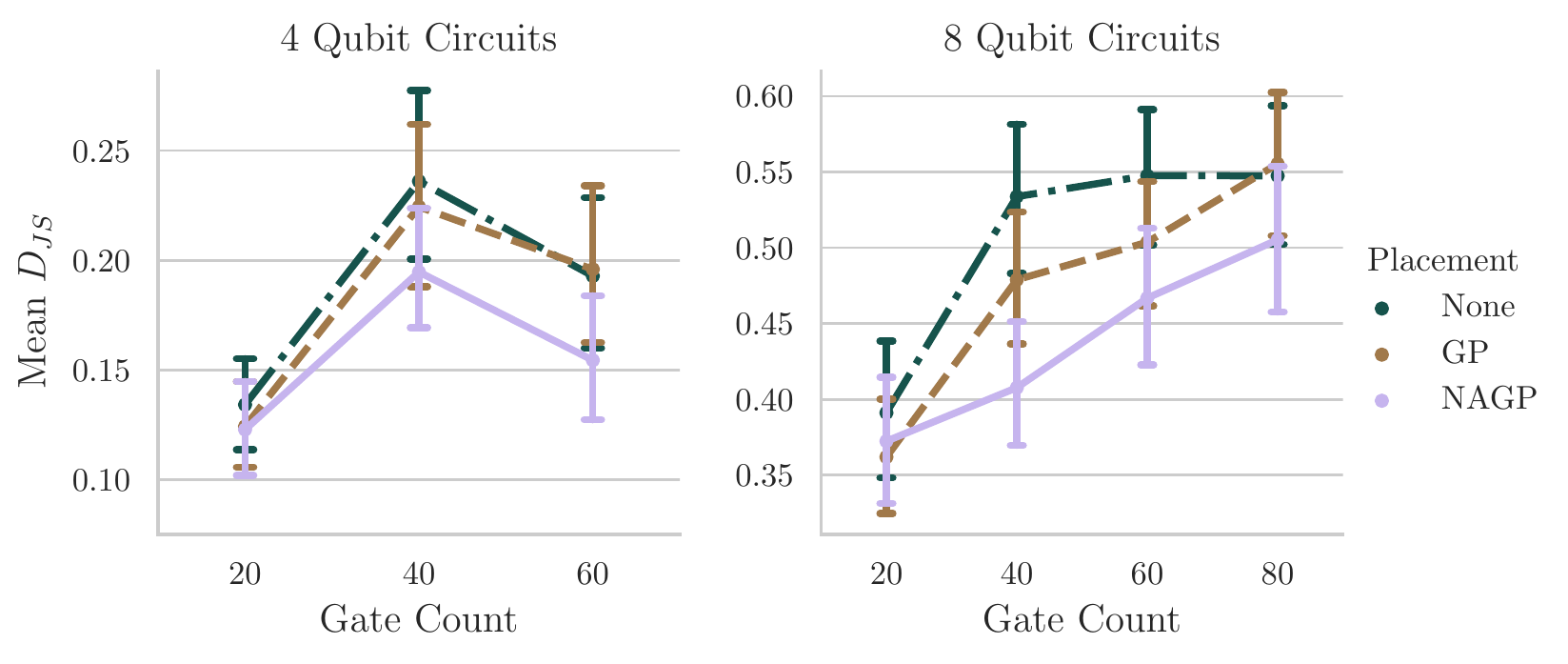}
  \caption{Experimental comparison of placement methods. ``None'' refers to the case of no initial placement of qubits, ``GP'' to partial placement via graph placement, and ``NAGP'' to noise-aware graph placement. Each data point corresponds to a mean Jensen--Shannon divergence \(D_\mathrm{JS}\) of measured distribution from ideal distribution, over the 90~random input circuit samples of the given size, compiled and executed on the device with the three different placement methods.}
  \label{fig:noisebench}
\end{figure}

Figure~\ref{fig:noisebench} plots the mean value of \(D_\mathrm{JS}\) over each
benchmark circuit set for the three placement methods, each benchmark set
parameterised by qubit number and gate count. In general, GP is seen to reduce
the mean \(D_\mathrm{JS}\) when compared to no initial placement; this can be
explained by GP reducing the number of error-prone two-qubit gates that need to
be added to map the circuit. We also see that scoring of these placements using
device-reported error information by NAGP is able to make further significant
reductions to \(D_\mathrm{JS}\), suggesting that such exploitation of device
heterogeneity is a worthwhile avenue of exploration for maximising near-term
device use.

Comparing 4-qubit and 8-qubit results, \(D_\mathrm{JS}\) means are higher in the
8-qubit case, as expected, as more qubits are entangled together in general and
so the system is more prone to error. In the 8-qubit case, \(D_\mathrm{JS}\) is
also seen to monotonically increase with gate count, also matching expectations.
The peak for \(D_\mathrm{JS}\) mean at 40~gates for 4~qubits, for all placement
methods, is unexpected and warrants further investigation.

\section{Conclusions and Future Work}
\label{sec:concl-future-work}

In this paper we have described CQC's compiler system \tket, with
particular emphasis on its transformation engine and qubit mapping
routine.  We showed that \tket offers significant improvement in terms
of gate count and gate depth over other comparable compiler systems
when evaluated on realistic quantum circuits and real quantum
architectures.  Further, for devices with heterogeneous gate and qubit
error rates \tket can use the component-level fidelity information to
appreciably improve overall device performance.  For NISQ-era quantum
computing such performance differences may be the margin between
success and failure.

The flexible design of \tket presents many possibilities for future
improvements.  Here we sketch three promising directions.

For all-to-all connectivity, the \texttt{PauliSimp} pass achieved
staggering depth reductions on the chemistry benchmark set, which is
not totally surprising because it was designed to exploit the
recurring structures found in UCCSD ans\"atze.  However there is a lot
of scope to improve this method, particularly if the Pauli Gadgets are
treated as multi-qubit gates and synthesised by the architecture-aware
phase of the compilation process, solving the problem mentioned in
Section~\ref{sec:end_to_end_results}. More generally we expect the use
of higher level ``big gates'', equipped with their own equational
theory, and tuned to particular algorithms or ans\"atze (for example
QAOA instances) to yield similar improvements.  This kind of
application-specific optimisations cannot be discovered by working
with random circuits and require real use cases.

Recent results on quantum volume \cite{Cross:2018aa} suggest that
available qubit numbers already surpass the gate fidelity by such a
margin that a large fraction of the qubits cannot effectively be
exploited.  This suggests that, in the near term at least, extremely
shallow circuits will be required.  One possible route to such depth
reduction is to exploit large numbers of ancillary qubits, combined
with techniques from measurement-based quantum computation
\cite{NIELSEN2006147} to effectively trade time for space.  The
\zxcalculus \cite{Coecke:2009aa} is already incorporated into \tket
and has proven an effective tool for MBQC calculations in the past
\cite{Duncan:2019aa,Duncan:2012uq,Duncan:2010aa}.

Finally, we will look at techniques to attack the noisiness of NISQ
devices head on.  Incorporating a very naive noise model into \tket's
qubit placement algorithm (Section~\ref{sec:noise-aware-plac}) made a
noticeable difference our results.  However it is well known that the
noise channels in real devices are much more complex and more
difficult to characterise
\cite{Harper:2019aa,Erhard:2019aa,Sung:2019aa}.  Incorporating better
analysis of the device errors into the compilation process, and
techniques to suppress and mitigate errors
\cite{Ball:2020aa,PhysRevA.94.052325,Kandala:2018ab} surely have a
role to play in the compilation process for NISQ devices for the
foreseeable future.

\clearpage
\section*{References}

\small
\bibliography{all}

\begin{thebibliography}{10}

\bibitem{Shor:PolyTimeFact:1997}
P.~W. Shor.
\newblock Polynomial-time algorithms for prime factorization and discrete
  logarithms on a quantum computer.
\newblock {\em SIAM J.Sci.Statist.Comput.}, 26(5), 1997.

\bibitem{Grover:1997qc}
Lov~K. Grover.
\newblock Quantum computers can search arbitrarily large databases by a single
  query.
\newblock {\em Phys. Rev. Lett.}, 79(23):4709--4712, 1997.

\bibitem{PhysRevLett.103.150502}
Aram~W. Harrow, Avinatan Hassidim, and Seth Lloyd.
\newblock Quantum algorithm for linear systems of equations.
\newblock {\em Phys. Rev. Lett.}, 103:150502, Oct 2009.

\bibitem{Georgescu:2013aa}
I.~M. Georgescu, S.~Ashhab, and Franco Nori.
\newblock Quantum simulation.
\newblock {\em Rev. Mod. Phys.}, 2013, arXiv:1308.6253.

\bibitem{10.1145/609784.609818}
Grace~Murray Hopper.
\newblock The education of a computer.
\newblock In {\em Proceedings of the 1952 ACM National Meeting (Pittsburgh)},
  ACM '52, pages 243--249, New York, NY, USA, 1952. Association for Computing
  Machinery.

\bibitem{Godbolt2019}
Matt Godbolt.
\newblock Optimizations in {C}++ compilers.
\newblock {\em Queue}, 17(5):69--100, October 2019.

\bibitem{Knill:Pseudocode:1996}
E.~Knill.
\newblock Conventions for quantum pseudocode.
\newblock Technical Report LAUR-96-2724, Los Alamos National Laboratory, 1996.

\bibitem{ibm:qiskit-aqua}
{IBM Research}.
\newblock Qiskit {A}qua.
\newblock \url{https://qiskit.org/aqua/}.

\bibitem{Bergholm:aa}
Ville Bergholm, Josh Izaac, Maria Schuld, Christian Gogolin, Carsten Blank,
  Keri McKiernan, and Nathan Killoran.
\newblock Pennylane: Automatic differentiation of hybrid quantum-classical
  computations.
\newblock {\em arXiv.org}, 2018, arXiv:1811.04968.

\bibitem{IBMQuantumExp}
{IBM Quantum Experience}.
\newblock \url{https://quantum-computing.ibm.com}.

\bibitem{rigetti-systems}
Rigetti {C}omputing.
\newblock \url{https://www.rigetti.com/systems}.

\bibitem{google-systems}
Google {Q}uantum.
\newblock \url{https://research.google/teams/applied-science/quantum/}.

\bibitem{Wright:2019aa}
K.~Wright, K.~M. Beck, S.~Debnath, J.~M. Amini, Y.~Nam, N.~Grzesiak, J.~S.
  Chen, N.~C. Pisenti, M.~Chmielewski, C.~Collins, K.~M. Hudek, J.~Mizrahi,
  J.~D. Wong-Campos, S.~Allen, J.~Apisdorf, P.~Solomon, M.~Williams, A.~M.
  Ducore, A.~Blinov, S.~M. Kreikemeier, V.~Chaplin, M.~Keesan, C.~Monroe, and
  J.~Kim.
\newblock Benchmarking an 11-qubit quantum computer.
\newblock {\em Nature Communications}, 10(1):5464, 2019.

\bibitem{honeywellquantumsystems}
Honeywell quantum systems.
\newblock \url{https://www.honeywell.com/en-us/company/quantum}, 2020.

\bibitem{xanadu-hardware}
Xanadu {H}ardware.
\newblock \url{https://www.xanadu.ai/hardware/}.

\bibitem{Rudolph:2016aa}
Terry Rudolph.
\newblock Why i am optimistic about the silicon-photonic route to quantum
  computing.
\newblock {\em arXiv preprint}, 2016, arXiv:1607.08535.

\bibitem{Preskill2018quantumcomputingin}
John Preskill.
\newblock Quantum {C}omputing in the {NISQ} era and beyond.
\newblock {\em {Quantum}}, 2:79, August 2018.

\bibitem{NieChu:QuantComp:2000}
M.~A. Nielsen and I.~L. Chuang.
\newblock {\em Quantum Computation and Quantum Information}.
\newblock Cambridge University Press, 2000.

\bibitem{O_Brien_2019}
Thomas~E O'Brien, Brian Tarasinski, and Barbara~M Terhal.
\newblock Quantum phase estimation of multiple eigenvalues for small-scale
  (noisy) experiments.
\newblock {\em New Journal of Physics}, 21(2):023022, feb 2019,
  arXiv:1809.09697.

\bibitem{Peruzzo:2014aa}
Alberto Peruzzo, Jarrod McClean, Peter Shadbolt, Man-Hong Yung, Xiao-Qi Zhou,
  Peter~J. Love, Al{\'a}n Aspuru-Guzik, and Jeremy~L. O'Brien.
\newblock A variational eigenvalue solver on a photonic quantum processor.
\newblock {\em Nature Communications}, 5, 07 2014, arXiv:1304.3061.

\bibitem{Farhi:2014aa}
Edward Farhi, Jeffrey Goldstone, and Sam Gutmann.
\newblock A quantum approximate optimization algorithm.
\newblock {\em arXiv.org}, 2014, arXiv:1411.4028.

\bibitem{Kandala:2018ab}
Abhinav Kandala, Kristan Temme, Antonio~D. Corcoles, Antonio Mezzacapo,
  Jerry~M. Chow, and Jay~M. Gambetta.
\newblock Error mitigation extends the computational reach of a noisy quantum
  processor.
\newblock {\em Nature}, 567:491--495, 2019, arxiv:1805.04492.

\bibitem{PhysRevA.94.052325}
Joel~J. Wallman and Joseph Emerson.
\newblock Noise tailoring for scalable quantum computation via randomized
  compiling.
\newblock {\em Phys. Rev. A}, 94:052325, Nov 2016, arXiv:1512.01098.

\bibitem{PhysRevLett.121.220502}
Bibek Pokharel, Namit Anand, Benjamin Fortman, and Daniel~A. Lidar.
\newblock Demonstration of fidelity improvement using dynamical decoupling with
  superconducting qubits.
\newblock {\em Phys. Rev. Lett.}, 121:220502, Nov 2018, arXiv:1807.08768.

\bibitem{Ball:2020aa}
Harrison Ball, Michael~J. Biercuk, Andre Carvalho, Rajib Chakravorty, Jiayin
  Chen, Leonardo~A. de~Castro, Steven Gore, David Hover, Michael Hush, Per~J.
  Liebermann, Robert Love, Kevin Nguyen, Viktor~S. Perunicic, Harry~J. Slatyer,
  Claire Edmunds, Virginia Frey, Cornelius Hempel, and Alistair Milne.
\newblock Software tools for quantum control: Improving quantum computer
  performance through noise and error suppression.
\newblock {\em arXiv.org}, 2020, arXiv:2001.04060.

\bibitem{Hner2016A-Software-Meth}
Thomas H\"aner, Damian~S. Steiger, Krysta Svore, and Matthias Troyer.
\newblock A software methodology for compiling quantum programs.
\newblock {\em arXiv.org}, (1604.01401), 2016, arXiv:1604.01401.

\bibitem{Alexander-S.-Green:2013fk}
Alexander~S. Green, Peter~LeFanu Lumsdaine, Neil~J. Ross, Peter Selinger, and
  Beno{\^\i}t Valiron.
\newblock Quipper: A scalable quantum programming language.
\newblock In {\em Programming language design and implementation (PLDI'13)},
  volume~48 of {\em ACM SIGPLAN Notices}, pages 333--342, 2013,
  arXiv:1304.3390.

\bibitem{JavadiAbhari20152}
Ali JavadiAbhari, Shruti Patil, Daniel Kudrow, Jeff Heckey, Alexey Lvov,
  Frederic~T. Chong, and Margaret Martonosi.
\newblock Scaffcc: Scalable compilation and analysis of quantum programs.
\newblock {\em Parallel Computing}, 45:2 -- 17, 2015, arXiv:1507.01902.
\newblock Computing Frontiers 2014: Best Papers.

\bibitem{Svore:2018:QES:3183895.3183901}
Krysta Svore, Alan Geller, Matthias Troyer, John Azariah, Christopher Granade,
  Bettina Heim, Vadym Kliuchnikov, Mariia Mykhailova, Andres Paz, and Martin
  Roetteler.
\newblock Q\#: Enabling scalable quantum computing and development with a
  high-level dsl.
\newblock In {\em Proceedings of the Real World Domain Specific Languages
  Workshop 2018}, RWDSL2018, pages 7:1--7:10, New York, NY, USA, 2018. ACM.

\bibitem{Killoran2019strawberryfields}
Nathan Killoran, Josh Izaac, Nicol{\'{a}}s Quesada, Ville Bergholm, Matthew
  Amy, and Christian Weedbrook.
\newblock Strawberry {F}ields: {A} {S}oftware {P}latform for {P}hotonic
  {Q}uantum {C}omputing.
\newblock {\em {Quantum}}, 3:129, Mar 2019.

\bibitem{forest}
{Rigetti Computing}.
\newblock Forest - rigetti.
\newblock \url{http://rigetti.com/forest}.

\bibitem{ibm:qiskit}
H{\'e}ctor Abraham, Ismail~Yunus Akhalwaya, Gadi Aleksandrowicz, Thomas
  Alexander, Gadi Alexandrowics, Eli Arbel, Abraham Asfaw, Carlos Azaustre,
  AzizNgoueya, Panagiotis Barkoutsos, George Barron, Luciano Bello, Yael
  Ben-Haim, Daniel Bevenius, Lev~S. Bishop, Samuel Bosch, David Bucher, CZ,
  Fran Cabrera, Padraic Calpin, Lauren Capelluto, Jorge Carballo, Gin{\'e}s
  Carrascal, Adrian Chen, Chun-Fu Chen, Richard Chen, Jerry~M. Chow, Christian
  Claus, Christian Clauss, Abigail~J. Cross, Andrew~W. Cross, Simon Cross, Juan
  Cruz-Benito, Cryoris, Chris Culver, Antonio~D. C{\'o}rcoles-Gonzales, Sean
  Dague, Matthieu Dartiailh, DavideFrr, Abd{\'o}n~Rodr{\'\i}guez Davila, Delton
  Ding, Eric Drechsler, Drew, Eugene Dumitrescu, Karel Dumon, Ivan Duran, Eric
  Eastman, Pieter Eendebak, Daniel Egger, Mark Everitt, Paco~Mart{\'\i}n
  Fern{\'a}ndez, Paco~Mart{\'\i}n Fern{\'a}ndez, Axel~Hern{\'a}ndez Ferrera,
  Albert Frisch, Andreas Fuhrer, MELVIN GEORGE, IAN GOULD, Julien Gacon, Gadi,
  Borja~Godoy Gago, Jay~M. Gambetta, Luis Garcia, Shelly Garion, Gawel-Kus,
  Juan Gomez-Mosquera, Salvador de~la Puente~Gonz{\'a}lez, Donny Greenberg,
  Dmitry Grinko, Wen Guan, John~A. Gunnels, Isabel Haide, Ikko Hamamura,
  Vojtech Havlicek, Joe Hellmers, {\L}ukasz Herok, Stefan Hillmich, Hiroshi
  Horii, Connor Howington, Shaohan Hu, Wei Hu, Haruki Imai, Takashi Imamichi,
  Kazuaki Ishizaki, Raban Iten, Toshinari Itoko, Ali Javadi-Abhari, Jessica,
  Kiran Johns, Naoki Kanazawa, Kang-Bae, Anton Karazeev, Paul Kassebaum,
  Knabberjoe, Arseny Kovyrshin, Vivek Krishnan, Kevin Krsulich, Gawel Kus, Ryan
  LaRose, Rapha{\"e}l Lambert, Joe Latone, Scott Lawrence, Dennis Liu, Peng
  Liu, Panagiotis Barkoutsos~ZRL Mac, Yunho Maeng, Aleksei Malyshev, Jakub
  Marecek, Manoel Marques, Dolph Mathews, Atsushi Matsuo, Douglas~T. McClure,
  Cameron McGarry, David McKay, Srujan Meesala, Antonio Mezzacapo, Rohit Midha,
  Zlatko Minev, Michael~Duane Mooring, Renier Morales, Niall Moran, Prakash
  Murali, Jan M{\"u}ggenburg, David Nadlinger, Giacomo Nannicini, Paul Nation,
  Yehuda Naveh, Nick-Singstock, Pradeep Niroula, Hassi Norlen, Lee~James
  O'Riordan, Oluwatobi Ogunbayo, Pauline Ollitrault, Steven Oud, Dan Padilha,
  Hanhee Paik, Simone Perriello, Anna Phan, Marco Pistoia, Alejandro
  Pozas-iKerstjens, Viktor Prutyanov, Daniel Puzzuoli, Jes{\'u}s P{\'e}rez,
  Quintiii, Rudy Raymond, Rafael Mart{\'\i}n-Cuevas Redondo, Max Reuter,
  Diego~M. Rodr{\'\i}guez, Mingi Ryu, Tharrmashastha SAPV, SamFerracin, Martin
  Sandberg, Ninad Sathaye, Bruno Schmitt, Chris Schnabel, Travis~L. Scholten,
  Eddie Schoute, Ismael~Faro Sertage, Nathan Shammah, Yunong Shi, Adenilton
  Silva, Yukio Siraichi, Iskandar Sitdikov, Seyon Sivarajah, John~A. Smolin,
  Mathias Soeken, Dominik Steenken, Matt Stypulkoski, Hitomi Takahashi, Charles
  Taylor, Pete Taylour, Soolu Thomas, Mathieu Tillet, Maddy Tod, Enrique de~la
  Torre, Kenso Trabing, Matthew Treinish, TrishaPe, Wes Turner, Yotam Vaknin,
  Carmen~Recio Valcarce, Francois Varchon, Desiree Vogt-Lee, Christophe
  Vuillot, James Weaver, Rafal Wieczorek, Jonathan~A. Wildstrom, Robert Wille,
  Erick Winston, Jack~J. Woehr, Stefan Woerner, Ryan Woo, Christopher~J. Wood,
  Ryan Wood, Stephen Wood, James Wootton, Daniyar Yeralin, Jessie Yu,
  Christopher Zachow, Laura Zdanski, Zoufalc, anedumla, azulehner, bcamorrison,
  brandhsn, chlorophyll zz, dime10, drholmie, elfrocampeador, faisaldebouni,
  fanizzamarco, gruu, kanejess, klinvill, kurarrr, lerongil, ma5x, merav
  aharoni, mrossinek, neupat, ordmoj, sethmerkel, strickroman, sumitpuri,
  tigerjack, toural, willhbang, yang.luh, and yotamvakninibm.
\newblock Qiskit: An open-source framework for quantum computing, 2019.

\bibitem{Steiger2016ProjectQ:-An-Op}
Damian~S. Steiger, Thomas H\"aner, and Matthias Troyer.
\newblock {ProjectQ}: An open source software framework for quantum computing.
\newblock {\em arXiv.org}, (1612.08091), 2016, arXiv:1612.08091.

\bibitem{cirq}
{The Cirq Developers}.
\newblock Cirq: A python library for nisq circuits.
\newblock \url{https://cirq.readthedocs.io/en/stable/}.

\bibitem{XACC}
Alexander McCaskey, Eugene Dumitrescu, Dmitry Lyakh, M.~Chen, W.~Feng, and
  Travis Humble.
\newblock A language and hardware independent approach to quantum--classical
  computing.
\newblock {\em SoftwareX}, 7:245--254, Jan 2018.

\bibitem{Murali:2019aa}
Prakash Murali, Norbert~Matthias Linke, Margaret Martonosi, Ali~Javadi Abhari,
  Nhung~Hong Nguyen, and Cinthia~Huerta Alderete.
\newblock Full-stack, real-system quantum computer studies: Architectural
  comparisons and design insights.
\newblock In {\em Proc. ICSA 2019}, 2019, arXiv:1905.11349.

\bibitem{robert_s_smith_2020_3677537}
Robert~S Smith, Eric~C Peterson, Erik~J Davis, and Mark~G Skilbeck.
\newblock quilc: An optimizing quil compiler, February 2020.

\bibitem{Nam:2018aa}
Yunseong Nam, Neil~J. Ross, Yuan Su, Andrew~M. Childs, and Dmitri Maslov.
\newblock Automated optimization of large quantum circuits with continuous
  parameters.
\newblock {\em npj Quantum Information}, 4(1):23, 2018.

\bibitem{Venturelli2019QuantumCC}
Davide Venturelli, Minh~Binh Do, Bryan O'Gorman, Jeremy Frank, Eleanor~G.
  Rieffel, Kyle E.~C. Booth, Th{\`a}nh~Nhut Nguyen, P.~P.~S. Narayan, and Sasha
  Nanda.
\newblock Quantum circuit compilation : An emerging application for automated
  reasoning.
\newblock In {\em 12th International Scheduling and Planning Application
  Workshop (SPARK)}, 2019.

\bibitem{Murali:2019ab}
Prakash Murali, Jonathan~M. Baker, Ali~Javadi Abhari, Frederic~T. Chong, and
  Margaret Martonosi.
\newblock Noise-adaptive compiler mappings for noisy intermediate-scale quantum
  computers.
\newblock {\em arXiv.org}, 2019, arXiv:1901.11054.

\bibitem{murali2020software}
Prakash Murali, David~C. McKay, Margaret Martonosi, and Ali Javadi-Abhari.
\newblock Software mitigation of crosstalk on noisy intermediate-scale quantum
  computers.
\newblock {\em arXiv}, 2020, 2001.02826.

\bibitem{peterson2019fixeddepth}
Eric~C. Peterson, Gavin~E. Crooks, and Robert~S. Smith.
\newblock Fixed-depth two-qubit circuits and the monodromy polytope.
\newblock {\em arXiv.org}, 2019, 1904.10541.

\bibitem{PhysRevA.95.042318}
Nelson Leung, Mohamed Abdelhafez, Jens Koch, and David Schuster.
\newblock Speedup for quantum optimal control from automatic differentiation
  based on graphics processing units.
\newblock {\em Phys. Rev. A}, 95:042318, Apr 2017, arXiv:1612.04929.

\bibitem{Gokhale:2019:PCV:3352460.3358313}
Pranav Gokhale, Yongshan Ding, Thomas Propson, Christopher Winkler, Nelson
  Leung, Yunong Shi, David~I. Schuster, Henry Hoffmann, and Frederic~T. Chong.
\newblock Partial compilation of variational algorithms for noisy
  intermediate-scale quantum machines.
\newblock In {\em Proceedings of the 52Nd Annual IEEE/ACM International
  Symposium on Microarchitecture}, MICRO '52, pages 266--278, New York, NY,
  USA, 2019. ACM, arXiv:1909.07522.

\bibitem{llvm:website}
The {LLVM} compiler infrastructure.
\newblock http://www.llvm.org.

\bibitem{Cross2017Open-Quantum-As}
Andrew~W. Cross, Lev~S. Bishop, John~A. Smolin, and Jay~M. Gambetta.
\newblock Open quantum assembly language.
\newblock
  \url{https://github.com/IBM/qiskit-openqasm/blob/master/spec/qasm2.pdf},
  January 2017.

\bibitem{Kissinger:2019ab}
Aleks Kissinger and John van~de Wetering.
\newblock Pyzx: Large scale automated diagrammatic reasoning.
\newblock {\em arXiv.org}, 2019, arXiv:1904.04735.

\bibitem{McClean:2017aa}
Jarrod~R. McClean, Kevin~J. Sung, Ian~D. Kivlichan, Yudong Cao, Chengyu Dai,
  E.~Schuyler Fried, Craig Gidney, Brendan Gimby, Pranav Gokhale, Thomas
  H{\"a}ner, Tarini Hardikar, Vojt{\v e}ch Havl{\'\i}{\v c}ek, Oscar Higgott,
  Cupjin Huang, Josh Izaac, Zhang Jiang, Xinle Liu, Sam McArdle, Matthew
  Neeley, Thomas O'Brien, Bryan O'Gorman, Isil Ozfidan, Maxwell~D. Radin,
  Jhonathan Romero, Nicholas Rubin, Nicolas P.~D. Sawaya, Kanav Setia, Sukin
  Sim, Damian~S. Steiger, Mark Steudtner, Qiming Sun, Wei Sun, Daochen Wang,
  Fang Zhang, and Ryan Babbush.
\newblock Openfermion: The electronic structure package for quantum computers.
\newblock {\em arXiv.org}, 2017, arXiv:1710.07629.

\bibitem{spam}
Mingyu {Sun} and Michael~R. {Geller}.
\newblock {Efficient characterization of correlated SPAM errors}.
\newblock {\em ArXiv.org}, 2018, arXiv:1810.10523.

\bibitem{Wootters1982A-single-quantu}
W.~Wootters and W.~Zurek.
\newblock A single quantum cannot be cloned.
\newblock {\em Nature}, 299:802--803, 1982.

\bibitem{Pati2000Impossibility-o}
A.K. Pati and S.~L. Braunstein.
\newblock Impossibility of deleting an unknown quantum state.
\newblock {\em Nature}, 404:164--165, 2000.

\bibitem{baez:2019aa}
John~C. Baez and Kenny Courser.
\newblock Structured cospans.
\newblock {\em arXiv.org}, 2019, arXiv:1911.04630.

\bibitem{IEEE:2019aa}
Ieee standard for floating-point arithmetic.
\newblock {\em IEEE Std 754-2019 (Revision of IEEE 754-2008)}, pages 1--84,
  2019.

\bibitem{maslov2016basic}
Dmitri Maslov.
\newblock Basic circuit compilation techniques for an ion-trap quantum machine.
\newblock {\em New Journal of Physics}, 19(023035), 2017, arXiv:1603.07678.

\bibitem{PhysRevA.88.052307}
Vadym Kliuchnikov and Dmitri Maslov.
\newblock Optimization of clifford circuits.
\newblock {\em Phys. Rev. A}, 88:052307, Nov 2013.

\bibitem{Ehrig:2006ab}
Hartmut Ehrig, Karsten Ehrig, Ulrike Prange, and Gabriele Taentzer.
\newblock {\em Fundamentals of Algebraic Graph Transformation}.
\newblock Monographs in Theoretical Computer Science. Springer Berlin
  Heidelberg, 2006.

\bibitem{contract_2001}
Richard Mitchell, Jim McKim, and Bertrand Meyer.
\newblock {\em Design by Contract, by Example}.
\newblock 0201634600. Addison Wesley Longman Publishing Co., Inc., USA, 2001.

\bibitem{Cross:2018aa}
Andrew~W. Cross, Lev~S. Bishop, Sarah Sheldon, Paul~D. Nation, and Jay~M.
  Gambetta.
\newblock Validating quantum computers using randomized model circuits.
\newblock {\em arXiv.org}, 2018, arXiv:1811.12926.

\bibitem{Blume-Kohout:2019ab}
Robin Blume-Kohout and Kevin~C. Young.
\newblock A volumetric framework for quantum computer benchmarks.
\newblock {\em arXiv preprint}, 2019, arXiv:1904.05546.

\bibitem{Erhard:2019aa}
Alexander Erhard, Joel~James Wallman, Lukas Postler, Michael Meth, Roman
  Stricker, Esteban~Adrian Martinez, Philipp Schindler, Thomas Monz, Joseph
  Emerson, and Rainer Blatt.
\newblock Characterizing large-scale quantum computers via cycle benchmarking.
\newblock {\em arXiv.org}, 2019, arXiv:1902.08543.

\bibitem{Arute:2019aa}
Frank Arute, Kunal Arya, Ryan Babbush, Dave Bacon, Joseph~C. Bardin, Rami
  Barends, Rupak Biswas, Sergio Boixo, Fernando G. S.~L. Brandao, David~A.
  Buell, Brian Burkett, Yu~Chen, Zijun Chen, Ben Chiaro, Roberto Collins,
  William Courtney, Andrew Dunsworth, Edward Farhi, Brooks Foxen, Austin
  Fowler, Craig Gidney, Marissa Giustina, Rob Graff, Keith Guerin, Steve
  Habegger, Matthew~P. Harrigan, Michael~J. Hartmann, Alan Ho, Markus Hoffmann,
  Trent Huang, Travis~S. Humble, Sergei~V. Isakov, Evan Jeffrey, Zhang Jiang,
  Dvir Kafri, Kostyantyn Kechedzhi, Julian Kelly, Paul~V. Klimov, Sergey Knysh,
  Alexander Korotkov, Fedor Kostritsa, David Landhuis, Mike Lindmark, Erik
  Lucero, Dmitry Lyakh, Salvatore Mandr{\`a}, Jarrod~R. McClean, Matthew
  McEwen, Anthony Megrant, Xiao Mi, Kristel Michielsen, Masoud Mohseni, Josh
  Mutus, Ofer Naaman, Matthew Neeley, Charles Neill, Murphy~Yuezhen Niu, Eric
  Ostby, Andre Petukhov, John~C. Platt, Chris Quintana, Eleanor~G. Rieffel,
  Pedram Roushan, Nicholas~C. Rubin, Daniel Sank, Kevin~J. Satzinger, Vadim
  Smelyanskiy, Kevin~J. Sung, Matthew~D. Trevithick, Amit Vainsencher, Benjamin
  Villalonga, Theodore White, Z.~Jamie Yao, Ping Yeh, Adam Zalcman, Hartmut
  Neven, and John~M. Martinis.
\newblock Quantum supremacy using a programmable superconducting processor.
\newblock {\em Nature}, 574(7779):505--510, 2019.

\bibitem{Gottesman:1999to}
Daniel Gottesman.
\newblock The heisenberg representation of quantum computers.
\newblock In S.~P. Corney, R.~Delbourgo, and P.~D. Jarvis, editors, {\em
  Proceedings of the XXII International Colloquium on Group Theoretical Methods
  in Physics}, pages 32--43. International Press, 1999, arXiv:quant-ph/9807006.

\bibitem{Scott-Aaronson:2004yf}
S.~Aaronson and D~Gottesman.
\newblock Improved simulation of stabilizer circuits.
\newblock {\em Phys. Rev. A}, 70(052328), 2004, arXiv:quant-ph/0406196v5.

\bibitem{amy2016finite}
Matthew Amy, Jianxin Chen, and Neil~J Ross.
\newblock A finite presentation of cnot-dihedral operators.
\newblock {\em arXiv preprint arXiv:1701.00140}, 2016.

\bibitem{EPTCS287.5}
Andrew Fagan and Ross Duncan.
\newblock Optimising clifford circuits with quantomatic.
\newblock In Peter Selinger and Giulio Chiribella, editors, {\em {\rm
  Proceedings of the 15th International Conference on} Quantum Physics and
  Logic, {\rm Halifax, Canada, 3-7th June 2018}}, volume 287 of {\em Electronic
  Proceedings in Theoretical Computer Science}, pages 85--105. Open Publishing
  Association, 2019, arXiv:1901.10114.

\bibitem{selinger2013generators}
Peter Selinger.
\newblock Generators and relations for n-qubit clifford operators.
\newblock {\em Logical Methods in Computer Science}, 11(2:10), 2015,
  arXiv:1310.6813.

\bibitem{khaneja2001cartan}
Navin Khaneja and Steffen~J. Glaser.
\newblock Cartan decomposition of su(2n) and control of spin systems.
\newblock {\em Chemical Physics}, 267(1-3):11--23, 2001.

\bibitem{blaauboer2008analytical}
M~Blaauboer and RL~De~Visser.
\newblock An analytical decomposition protocol for optimal implementation of
  two-qubit entangling gates.
\newblock {\em Journal of Physics A: Mathematical and Theoretical},
  41(39):395307, 2008.

\bibitem{vidal2004universal}
Guifre Vidal and Christopher~M Dawson.
\newblock Universal quantum circuit for two-qubit transformations with three
  controlled-not gates.
\newblock {\em Physical Review A}, 69(1):010301, 2004.

\bibitem{Cowtan:2019aa}
Alexander Cowtan, Silas Dilkes, Ross Duncan, Will Simmons, and Seyon Sivarajah.
\newblock Phase gadget synthesis for shallow circuits.
\newblock In {\em Proceedings of QPL2019 (to appear)}, 2019, arXiv:1906.01734.

\bibitem{Coecke:2009aa}
Bob Coecke and Ross Duncan.
\newblock Interacting quantum observables: Categorical algebra and
  diagrammatics.
\newblock {\em New J. Phys}, 13(043016), 2011, arXiv:0906.4725.

\bibitem{Amy2014Polynomial-Time}
Matthew Amy, Dmitri Maslov, and Michele Mosca.
\newblock Polynomial-time t-depth optimization of clifford+t circuits via
  matroid partitioning.
\newblock {\em IEEE Transactions on Computer-Aided Design of Integrated
  Circuits and Systems}, 33(10):1476--1489, 2014.

\bibitem{Duncan:2019aa}
Ross Duncan, Aleks Kissinger, Simon Perdrix, and John van~de Wetering.
\newblock Graph-theoretic simplification of quantum circuits with the
  zx-calculus.
\newblock {\em arXiv.org}, 2019.

\bibitem{Kissinger:2019aa}
Aleks Kissinger and John van~de Wetering.
\newblock Reducing t-count with the zx-calculus.
\newblock {\em arXiv.org}, 2019, arXiv:1903.10477.

\bibitem{Maslov2017Shorter-stabili}
Dmitri Maslov and Martin Roetteler.
\newblock Shorter stabilizer circuits via bruhat decomposition and quantum
  circuit transformations.
\newblock {\em IEEE Transactions on Information Theory}, (7):4729--4738, 2017,
  arXiv:1705.09176.

\bibitem{Maslov2005Toffoli-network}
D.~Maslov, G.W. Dueck, and D.M. Miller.
\newblock Toffoli network synthesis with templates.
\newblock {\em Computer-Aided Design of Integrated Circuits and Systems, IEEE
  Transactions}, 24(6):807--817, June 2005.

\bibitem{Alexander-Cowtan:2019aa}
Alexander Cowtan, Silas Dilkes, Ross Duncan, Alexandre Krajenbrink, Will
  Simmons, and Seyon Sivarajah.
\newblock On the qubit routing problem.
\newblock In Wim van Dam and Laura Mancinska, editors, {\em 14th Conference on
  the Theory of Quantum Computation, Communication and Cryptography (TQC
  2019)}, volume 135 of {\em Leibniz International Proceedings in Informatics
  (LIPIcs)}, pages 5:1--5:32, 2019.

\bibitem{Childs:2019aa}
Andrew~M. Childs, Eddie Schoute, and Cem~M. Unsal.
\newblock Circuit transformations for quantum architectures.
\newblock In Wim van Dam and Laura Mancinska, editors, {\em 14th Conference on
  the Theory of Quantum Computation, Communication and Cryptography (TQC
  2019)}, volume 135 of {\em Leibniz International Proceedings in Informatics
  (LIPIcs)}, pages 3:1--3:24, 2019, arXiv:1902.09102.

\bibitem{Zulehner:2018aa}
Alwin Zulehner and Robert Wille.
\newblock Compiling su(4) quantum circuits to ibm qx architectures.
\newblock {\em arXiv.org}, 2018, arXiv:1808.05661.

\bibitem{Zulehner:2017aa}
Alwin Zulehner, Alexandru Paler, and Robert Wille.
\newblock An efficient methodology for mapping quantum circuits to the ibm qx
  architectures.
\newblock {\em arXiv.org}, 2017, arXiv:1712.04722.

\bibitem{Tannu:2018aa}
Swamit~S. Tannu and Moinuddin K.Qureshi.
\newblock A case for variability-aware policies for nisq-era quantum computers.
\newblock {\em arXiv.org}, 2018, arXiv:1805.10224.

\bibitem{PhysRevA.95.042308}
Jarrod~R. McClean, Mollie~E. Kimchi-Schwartz, Jonathan Carter, and Wibe~A.
  de~Jong.
\newblock Hybrid quantum-classical hierarchy for mitigation of decoherence and
  determination of excited states.
\newblock {\em Phys. Rev. A}, 95:042308, Apr 2017, arXiv:1603.05681.

\bibitem{Romero_2018}
Jonathan Romero, Ryan Babbush, Jarrod~R McClean, Cornelius Hempel, Peter~J
  Love, and Al{\'{a}}n Aspuru-Guzik.
\newblock Strategies for quantum computing molecular energies using the unitary
  coupled cluster ansatz.
\newblock {\em Quantum Science and Technology}, 4(1):014008, oct 2018.

\bibitem{Steudtner_2018}
Mark Steudtner and Stephanie Wehner.
\newblock Fermion-to-qubit mappings with varying resource requirements for
  quantum simulation.
\newblock {\em New Journal of Physics}, 20(6):063010, jun 2018.

\bibitem{Childs:2018aa}
Andrew~M. Childs, Dmitri Maslov, Yunseong Nam, Neil~J. Ross, and Yuan Su.
\newblock Toward the first quantum simulation with quantum speedup.
\newblock {\em Proceedings of the National Academy of Sciences}, 115(38):9456,
  09 2018.

\bibitem{QuantumComputingReport}
Doug Finke.
\newblock
  https://quantumcomputingreport.com/news/ibm-opens-new-quantum-data-center-introduces-new-53-qubit-machine/.
\newblock
  https://quantumcomputingreport.com/news/ibm-opens-new-quantum-data-center-introduces-new-53-qubit-machine/.

\bibitem{rigetti_blog}
Chad Rigetti.
\newblock
  https://medium.com/rigetti/the-rigetti-128-qubit-chip-and-what-it-means-for-quantum-df757d1b71ea.
\newblock
  https://medium.com/rigetti/the-rigetti-128-qubit-chip-and-what-it-means-for-quantum-df757d1b71ea.

\bibitem{ibmq}
{IBM} {Q}.
\newblock \url{https://www.ibm.com/quantum-computing/}.

\bibitem{Flammia:1346875}
Steven~T. Flammia and Yi-Kai Liu.
\newblock Direct fidelity estimation from few pauli measurements.
\newblock {\em Phys. Rev. Lett.}, 106:230501, Jun 2011, arXiv:1104.4695.

\bibitem{Nishio:2019aa}
Shin Nishio, Yulu Pan, Takahiko Satoh, Hideharu Amano, and Rodney~Van Meter.
\newblock Extracting success from ibm's 20-qubit machines using error-aware
  compilation.
\newblock {\em arXiv.org}, 2019.

\bibitem{PhysRevA.72.052310}
A.~P. Majtey, P.~W. Lamberti, and D.~P. Prato.
\newblock Jensen-shannon divergence as a measure of distinguishability between
  mixed quantum states.
\newblock {\em Phys. Rev. A}, 72:052310, Nov 2005.

\bibitem{NIELSEN2006147}
Michael~A. Nielsen.
\newblock Cluster-state quantum computation.
\newblock {\em Reports on Mathematical Physics}, 57(1):147 -- 161, 2006,
  arXiv:quant-ph/0504097.

\bibitem{Duncan:2012uq}
Ross Duncan.
\newblock A graphical approach to measurement-based quantum computing.
\newblock In Chris Heunen, Mehrnoosh Sadrzadeh, and Edward Grefenstette,
  editors, {\em Quantum Physics and Linguistics: A Compositional, Diagrammatic
  Discourse}, chapter~3. Oxford University Press, 2013, arxiv:1203.6242.

\bibitem{Duncan:2010aa}
R.~Duncan and S.~Perdrix.
\newblock Rewriting measurement-based quantum computations with generalised
  flow.
\newblock In S.~Abramsky, C.~Gavoille, C~Kirchner, F.~Meyer auf~der Heide, and
  P.~G. Spirakis, editors, {\em Automata, Languages and Programming, 37th
  International Colloquium, ICALP 2010, Proceedings Part II}, volume 6199 of
  {\em Lecture Notes in Computer Science}, pages 285--296. Springer, 2010.

\bibitem{Harper:2019aa}
Robin Harper, Steven~T. Flammia, and Joel~J. Wallman.
\newblock Efficient learning of quantum noise.
\newblock {\em arXiv.org}, 2019, arXiv:1907.13022.

\bibitem{Sung:2019aa}
Youngkyu Sung, F{\'e}lix Beaudoin, Leigh~M. Norris, Fei Yan, David~K. Kim,
  Jack~Y. Qiu, Uwe von L{\"u}pke, Jonilyn~L. Yoder, Terry~P. Orlando, Simon
  Gustavsson, Lorenza Viola, and William~D. Oliver.
\newblock Non-gaussian noise spectroscopy with a superconducting qubit sensor.
\newblock {\em Nature Communications}, 10(1):3715, 2019, arXiv:1903.01043.

\end{thebibliography}


\end{document}